\documentclass{vldb}
\usepackage{graphicx}
\usepackage{libertine}
\usepackage{epigraph}
\usepackage{balance}  % for  \balance command ON LAST PAGE  (only there!)
\usepackage{xspace}
\usepackage{textcomp}
\usepackage{makecell} % required for \thead
\usepackage{multirow}
\usepackage{subfig}
\usepackage{framed}
\usepackage{mathrsfs} % for math fonts
\usepackage{enumitem}
\usepackage{boldline}
\usepackage[dvipsnames]{xcolor}
\usepackage{booktabs}
\usepackage[urlcolor=blue,hidelinks]{hyperref}
\usepackage[labelfont=bf]{caption}

\newcommand{\gdpr}{$\mathscr{G}$}

\newcommand{\sref}[1]{Section-\ref{#1}}
\newcommand{\vheading}[1]{\vspace{0.05in}\noindent\textbf{#1}}

\newcommand{\etc}{\textit{etc.}\xspace}
\newcommand{\myx}{$\times$\xspace}

\vldbTitle{Understanding and Benchmarking the Impact of GDPR on Database Systems}
\vldbAuthors{Supreeth Shastri, Vinay Banakar, Melissa Wasserman, Arun Kumar, and Vijay Chidambaram}
\vldbDOI{https://doi.org/10.14778/3384345.3384354}
\vldbVolume{13}
\vldbNumber{7}
\vldbYear{2020}

\begin{document}

% ****************** TITLE ****************************************

\title{Understanding and Benchmarking\\[1mm] the Impact of GDPR on Database Systems}

% possible, but not really needed or used for PVLDB:
%\subtitle{\large (Experiments and Analysis paper)}
%\titlenote{A full version of this paper is available as\textit{Author's Guide to Preparing ACM SIG Proceedings Using \LaTeX$2_\epsilon$\ and BibTeX} at \texttt{www.acm.org/eaddress.htm}}}

% ****************** AUTHORS **************************************


\numberofauthors{5} %  in this sample file, there are a *total*

\author{
% e-mail address with \affaddr.
%
% 1st. author
\alignauthor
{\large Supreeth Shastri}\\[0.5mm]
       \affaddr{\normalsize UT Austin}\\
       \affaddr{\normalsize shastri@utexas.edu}\\
\alignauthor 
{\large Vinay Banakar}\\[0.5mm]
       \affaddr{\normalsize Hewlett Packard Enterprise}\\
       \affaddr{\normalsize vinay.s.banakar@gmail.com}\\
\alignauthor
{\large Melissa Wasserman}\\[0.5mm]
       \affaddr{\normalsize UT Austin, School of Law}\\
       \affaddr{\normalsize MWasserman@law.utexas.edu}\\
\and
\alignauthor 
{\large Arun Kumar}\\[0.5mm]
       \affaddr{\normalsize University of California, San Diego}\\
       \affaddr{\normalsize arunkk@eng.ucsd.edu}\\
\alignauthor 
{\large Vijay Chidambaram}\\[0.5mm]
       \affaddr{\normalsize UT Austin and VMware Research}\\
       \affaddr{\normalsize vijay@cs.utexas.edu}\\
%\vspace{5mm}
}

\setcounter{page}{1064}
\maketitle

\begin{abstract}
The General Data Protection Regulation (GDPR) provides new rights and
protections to European people concerning their personal data. We analyze GDPR
from a systems perspective, translating its legal articles into a set
of capabilities and characteristics that compliant systems must
support. Our analysis reveals the phenomenon of \emph{metadata
explosion}, wherein large quantities of metadata needs to be stored
along with the personal data to satisfy the GDPR requirements. Our
analysis also helps us identify new workloads that must be supported
under GDPR. We design and implement an open-source benchmark called
\emph{GDPRbench} that consists of workloads and metrics needed to
understand and assess personal-data processing database systems. To
gauge the readiness of modern database systems for GDPR, we follow
best practices and developer recommendations to modify Redis,
PostgreSQL, and a commercial database system to be GDPR compliant. Our
experiments demonstrate that the resulting GDPR-compliant systems
achieve poor performance on GPDR workloads, and that performance
scales poorly as the volume of personal data increases. We discuss the
real-world implications of these findings, and identify research
challenges towards making GDPR-compliance efficient in production
environments. We release all of our software artifacts and datasets at
{\color{blue}\url{http://www.gdprbench.org}}

\end{abstract}

{
%\vspace{1cm}
\section{Introduction}
\label{sec-introduction}
%\vspace{-2mm}
%\setlength{\epigraphwidth}{1.8in}
%\setlength{\epigraphrule}{0.1pt}
%\epigraph{\emph{``Measure what is measurable, and make measurable what is not so.''}}{Galileo Galilei}
%\vspace{-2mm}

The European Union enacted the General Data Protection Regulation
(GDPR)~\cite{gdpr-regulation} on May 25th 2018 to counter widespread
abuse of personal data. While at-scale monetization of personal data
has existed since the early dot-com days, the unprecedented rate at
which such data is getting compromised is a recent phenomenon.  To
counter this trend, GDPR declares the privacy and protection of
personal data as a fundamental right of all European people. It grants
several new \emph{rights to the EU consumers} including the right to
access, right to rectification, right to be forgotten, right to
object, and right to data portability. GDPR also assigns
\emph{responsibilities to companies} that collect and process personal
data. These include seeking explicit consent before using personal
data, notifying data breaches within 72 hours of discovery,
maintaining records of processing activities, \etc Failing to comply
with GDPR could result in hefty penalties: up to \texteuro20M or 4\%
of global revenue, whichever is higher. For instance, in January 2019,
Google was fined \texteuro50M for not obtaining customer consent in
their ads personalization~\cite{google-purpose-bundling}; in July
2019, British Airways was fined \textsterling184M for failing to
safeguard personal data of their customers~\cite{ba-gdpr-fine}.

Compliance with GDPR is
challenging for several reasons. First, GDPR's interpretation of
personal data is broad as it includes any information that relates to
a natural person, even if it did not uniquely identify that
person. For example, search terms sent to Google are covered under
GDPR.  This vastly increases the scope of data that comes under GDPR
purview. Second, several GDPR regulations are intentionally vague in
their technical specification to accommodate future advancements in
technologies. This causes confusion among developers of GDPR-compliant
systems. Finally, several GDPR requirements are fundamentally at odds
with the design principles and operating practices of modern computing
systems~\cite{gdpr-sins}.  It is no surprise that recent
estimates~\cite{iapp-prediction, gartner-prediction} peg the
compliance rate to be less than 50\%.

\vheading{Analyzing GDPR}. In this work, we aim to understand and
evaluate GDPR compliance of existing database systems. Our goal is not
to optimize these systems or to build new systems from scratch; instead, 
we follow public best practices and developer recommendations to
ensure existing systems are GDPR-compliant~\cite{gdpr-redis,
  postgresql}. We analyze GDPR and distill its articles into
capabilities and characteristics that database systems must
support. By design, the law allows multiple interpretations: we pick a
strict interpretation to reason about the worst-case performance costs
of GDPR compliance. For example, GDPR does not specify how soon after
a \emph{Right To Be Forgotten} request should the data be erased. We
resolve this ambiguity by requiring the deletion request to be
initiated (and possibly completed) within a few seconds. In contrast,
Google cloud, which claims GDPR compliance, informs that all deletions
will complete within 180 days of request~\cite{google-deletion}. We
believe that analyzing the impact of and benchmarking the overheads of
the worst-case performance (resulting from strict interpretation) is a
useful reference point for designers and administrators. However, in
practice, a company may adapt a relaxed interpretation that reduces
the cost and overhead of compliance. We make three key observations in
our analysis.

\begin{enumerate}[leftmargin=4mm, parsep=0.5mm, topsep=0.5mm]

\item {We identify and characterize the phenomenon of \emph{metadata
      explosion}, whereby every personal data item is associated with
    up to seven metadata properties (such as purpose, time-to-live,
    objections etc) that govern its behavior. By elevating each
    personal data item into an active entity that has its own set of
    rules, GDPR mandates that it could no longer be used as a fungible
    commodity. This is significant from a database standpoint as it
    severely impacts both the control- and data-path operations}. 

\item{We observe that GDPR's goal of \emph{data protection by design
      and by default} conflicts with the traditional system design
    goals of optimizing for performance, cost, and reliability. For
    example, in order to notify people affected by data breaches,
    a company may want to keep an audit trail of all accesses to 
    their personal data. From a datastore perspective, this turns every read
    operation into a read followed by a write.}

\item{Lastly, we identify that GDPR allows new forms of interactions
    with datastores. We discuss the characteristics of these novel
    \emph{GDPR queries} (which we organize into a new benchmark called
    \emph{GDPRBench}), and their implications for database systems.}

\end{enumerate}

\vheading{GDPRbench}. As our analysis reveals, GDPR significantly
affects the design and operation of datastores that hold personal
data. However, none of the existing benchmarks recognize the
abstraction of personal data: its characteristics, storage
restrictions, or interfacing requirements. We design and implement
\emph{GDPRbench}, a new open-source benchmark that represents the
functionalities of a datastore deployed by a company that collects and
processes personal data. The design of GDPRbench is informed by
painstaking analysis of the legal cases arising from GDPR from its
first year of roll-out. GDPRbench is composed of four core workloads:
\emph{Controller}, \emph{Customer}, \emph{Processor}, and
\emph{Regulator}; these core workloads are not captured by any
database benchmark available today. GDPRbench captures three
benchmarking metrics for each workload: correctness, completion time,
and storage overhead.

\vheading{Evaluating GDPR-Compliant DBMS}. Finally, we aim to gauge
the GDPR compliance of modern database systems. We take three
widely-used database systems, Redis~\cite{redis} (an in-memory NoSQL
store), PostgreSQL~\cite{postgresql} (an open-source RDBMS), and a
commercial enterprise-grade RDBMS, and modify them to be GDPR
compliant. We followed recommendations from the developers of these
tools~\cite{gdpr-redis, postgresql} in making them GDPR-compliant; the
goal was to make minimal changes, not to redesign the system internals
for efficient compliance. While all three were able to achieve GDPR
compliance with a small amount of effort, the resulting systems
experienced a performance slow down of 2-5\myx for traditional
workloads like YCSB, primarily due to monitoring and logging mandated
by GDPR. We evaluated the performance of these
databases against GDPR workloads using GDPRbench. We observe that GDPR
queries may result in large amounts of data records and metadata being
returned, resulting in significantly low throughput. Our analyses and
experiments identify several implications for administering
GDPR-compliant database systems in the real world.

\vheading{Limitations.} In this work, we have not tried to optimize
the performance of the GDPR-compliant systems we evaluate. We merely
followed developer recommendations to achieve GDPR compliance. We
realize that the resulting performance degradation could be further
reduced with GDPR-specific optimizations or by redesigning security
mechanisms. Thus, our work focuses on understanding GDPR compliance
resulting from retrofitting existing systems. Next, the design
of GDPRbench is guided by several factors including (i) our
interpretation of GDPR, (ii) court rulings and GDPR use-cases in the
real-world, and (iii) the three database systems that we
investigated. As such, we recognize that the current iteration of
GDPRbench is a snapshot in time, and may need to evolve as more
technical and legal use cases emerge.

\vheading{Summary of contributions.}
Our work lays the foundation for understanding and benchmarking the impact of GDPR on database systems. In particular, we make the following contributions:
\begin{itemize}[leftmargin=3mm, parsep=0.5mm, topsep=0.5mm]

\item{\textbf{GDPR Analysis:} Our work is one of the first to
    explore GDPR from a database systems perspective. We analyze GDPR 
    articles, both individually and collectively, to distill
    them into attributes and actions for database systems. In doing
    so, we (i) observe the phenomenon of metadata explosion, and (ii)
    identify new queries and workloads that personal data systems
    must support.}

\item{\textbf{Design and Implementation of GDPRbench:} To enable
    customers, companies and regulators interpret GDPR compliance in a
    rigorous and systematic way, we design an open-source benchmark
    named GDPRBench. In GDPRbench, we model the queries and workloads
    that datastores encounter in the real-world, and develop metrics
    to succinctly represent their behavior. We publicly release all of
    our software artifacts at
    {\color{blue}\url{http://www.gdprbench.org}}.}

\item{\textbf{Experimental Evaluation:} We discuss our effort into
    modifying Redis, PostgreSQL, and a commercial RDBMS to be
    GDPR-compliant. Our evaluation shows that GDPR compliance achieved
    by minimal changes via straightforward mechanisms results in
    significant performance degradation for traditional
    workloads. Using GDPRbench, we show the completion time and
    storage space overhead of these compliant systems against
    real-world GDPR workloads. Finally, we share our insights on
    deploying compliant systems in production environments,
    implications of scaling personal data, as well as the research
    challenges of efficient GDPR compliance.}
\end{itemize}

\section{Background}
\label{sec-background}

We begin with a primer on GDPR including its internal structure and its adoption challenges in the real world. 

\subsection{GDPR Overview}
The European parliament adopted GDPR on April 14th 2016, and made it an enforceable law in all its member states starting May 25th 2018. GDPR is written\footnote{\small{even the CS people in our team found it quite readable!}} as 99 \emph{articles} that describe its legal requirements, and 173 \emph{recitals} that provide additional context and clarifications to these articles. The articles (henceforth prefixed with \gdpr) could be grouped into five broad categories. \gdpr1-11 articles layout the definitions and principles of personal data processing; \gdpr12-23 establish the rights of the people; then \gdpr24-50 mandate the responsibilities of the data controllers and processors; the next 26 describe the role and tasks of supervisory authorities; and the remainder of them cover liabilities, penalties and specific situations. We expand on the three categories that concern systems storing personal data.

\vheading{Principles of data processing.}
GDPR establishes several core principles governing personal data. For example, \gdpr5 requires that data collection be for a specific purpose, be limited to what is necessary, stored only for a well defined duration, and protected against loss and destruction. \gdpr6 defines the lawful basis for processing, while \gdpr7 describes the role of consent. 

\vheading{Rights of data subjects.}
GDPR grants 12 explicit and excercisable rights to every data subject (a natural person whose personal data is collected). These rights are designed to keep people in loop throughout the lifecycle of their personal data. At the time of collection, people have the right to know the specific purpose and exact duration for which their data would be used (\gdpr13, 14). At any point, people can access their data (\gdpr15), rectify errors (\gdpr16), request erasure (\gdpr17), download or port their data to a third-party (\gdpr20), object to it being used for certain purposes (\gdpr21), or withdraw from automated decision-making (\gdpr22). In the rest of the paper, we use the terms, data subjects and customers, synonymously.

\vheading{Responsibilities of data controllers.}
The third group of articles outline the responsibilities of data controllers (entities that collect and utilize personal data) and data processors (entities that process personal data on behalf of a controller). To clarify, when Netflix runs their recommendation algorithm on Amazon's MapReduce platform, Netflix is the controller and Amazon, the processor. Key responsibilities include designing secure infrastructure (\gdpr24, 25), maintaining records of processing (\gdpr30), notifying data breaches within 72 hours (\gdpr33, 34), analyzing risks prior to processing large amounts of personal data (\gdpr35, 36) and controlling the location of data (\gdpr44). Additionally, the controllers should create interfaces for people to exercise their GDPR rights.

\subsection{GDPR from a Database Perspective}
GDPR defines four entities---controller, customer, processor, and regulator---that interact with the data store. Figure--\ref{fig:gdpr-dataflow} shows how three distinct types of data flows between the GDPR entities and data stores. The database that hosts personal data and its associated metadata (purpose, objections etc.,) is the focus of our work. We distinguish it from the other store that contains non-GDPR and derived data as the rules of GDPR do not apply to them. 

The controller is responsible for collection and timely deletion of personal data as well as managing its GDPR metadata throughout the lifecycle. The customers interact with the data store to exercise their GDPR rights. The processor uses the stored personal data to generate derived data and intelligence, which in turn powers the controller's businesses and services. Finally, the regulators interact with the datastores to investigate complaints and to ensure that rights and responsibilities are complied with. 

Our focus on datastores is motivated by the high proportion of GDPR articles that concern them. From out of the 99 GDPR articles, 31 govern the behavior of data storage systems. In contrast, only 11 describe requirements from compute and network infrastructure. This should not be surprising given that GDPR is more focused on the control-plane aspects of personal data (like collecting, securing, storing, moving, sharing, deleting etc.,) than the actual processing of it.

%--------------------------------------------------------------------
% GDPR roles and data types 
%--------------------------------------------------------------------
\begin{figure}[t]
\centering
\includegraphics[width=0.44\textwidth]{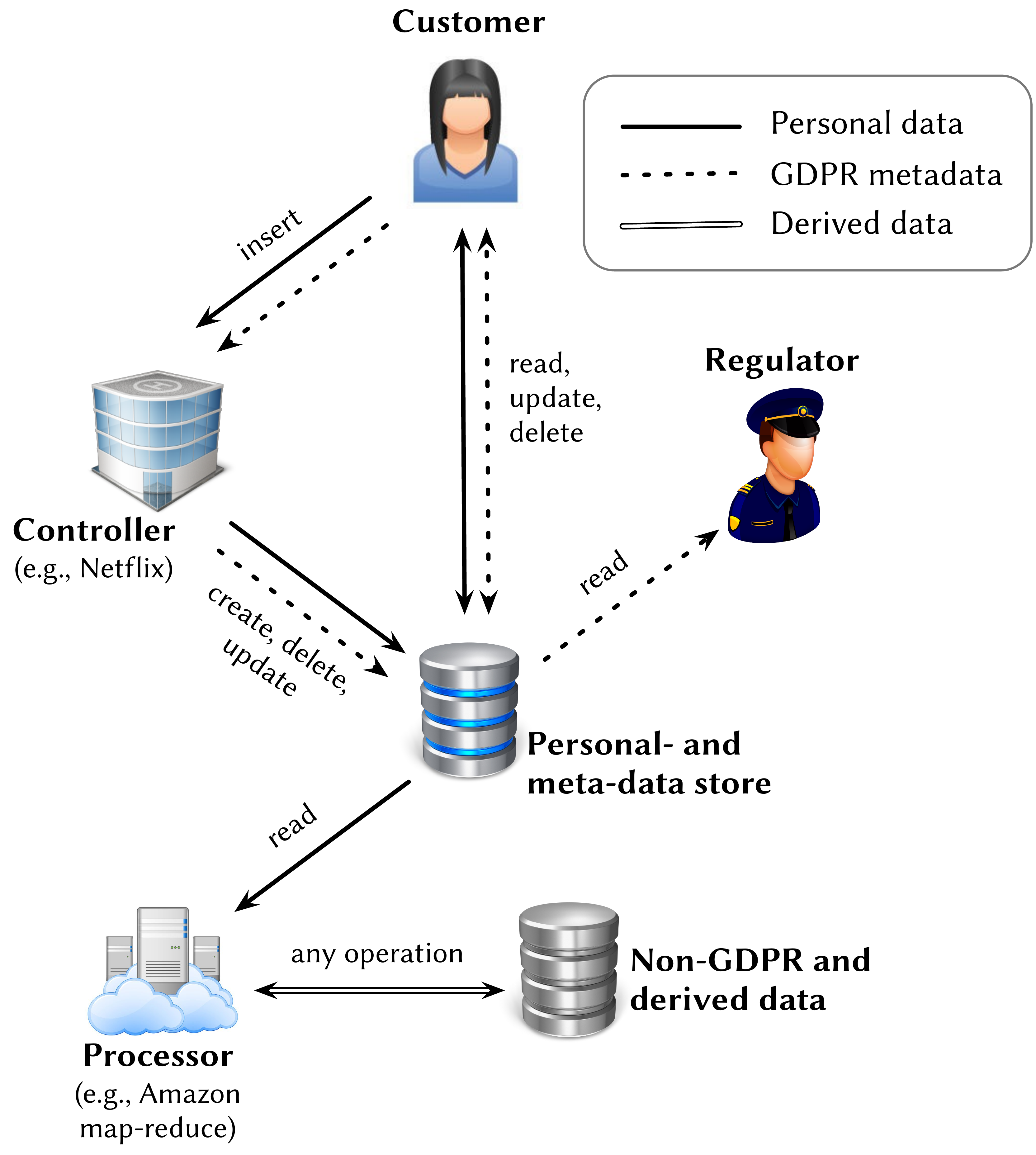}
\caption{\textbf {GDPR defines four roles and distinguishes between three types of data. The arrows out of a datastore indicate read-only access, while the arrows into it modify it. (1) The controller can collect, store, delete and update any personal- and GDPR-metadata, (2) A customer can read, update, or delete any personal data and GDPR-metadata that concerns them, (3) A processor reads personal data and produces derived data, and (4) Regulators access GDPR metadata to investigate customer complaints.}}
\label{fig:gdpr-dataflow}
\vspace{-4mm}
\end{figure}
%--------------------------------------------------------------------

\subsection{GDPR in the Wild}
The first year of GDPR has demonstrated both the need for and challenges of a comprehensive privacy law. On one hand, people have been exercising their newfound rights like the ability to download all the personal data that companies have amassed on them~\cite{gdpr-data-download}, and not been shy to report any shortcomings. In fact, the EU data protection board reports~\cite{edpb-8-months} 144,376 complaints from individuals and organizations in the first 12 months of GDPR. 

However, any attempt to regulate decade-long practices of commoditizing personal data is not without consequences. A number of companies like Instapaper, Klout, and Unroll.me voluntarily terminated their services in Europe to avoid the hassles of compliance. Like wise, most of the programmatic ad exchanges~\cite{gdpr-programmatic-ad-buy} of Europe were forced to shut down. This was triggered by Google and Facebook restricting access to their platforms to those ad exchanges that could not verify the legality of the personal data they possessed. But, several organizations could comply by making minor modifications to their business models. For example, media site \emph{USA Today} turned off all advertisements~\cite{usa-today-no-ads}, whereas \emph{the New York Times} stopped serving personalized ads~\cite{nytimes-no-ads}. 

As \gdpr28 precludes employing any data processor that does not comply with GDPR, the cloud providers have been swift in showcasing~\cite{aws-gdpr-ready, gce-gdpr-ready, azure-gdpr-ready} their compliance. However, given the monetary and technical challenges in redesigning the existing systems, the focus has unfortunately shifted to \emph{reactive security}. It is still an open question if services like Amazon Macie~\cite{amazon-macie}, which employs machine learning to automatically discover, monitor, and alert misuse of personal data on behalf of legacy cloud applications would survive the GDPR scrutiny.

Regulators have been active and vigilant as well: the number of ongoing and completed investigations in the first 9 months of GDPR is reported to be 206326. Regulators have already levied penalties on several companies including \texteuro50M on Google~\cite{google-purpose-bundling} for lacking a legal basis for their ads personalization, and \textsterling184M on British Airways~\cite{ba-gdpr-fine} for lacking security of processing. However, the clearest sign of GDPR's effectiveness is in the fact that regulators have received 89,271 voluntary data breach notifications from companies in the first 12 months of GDPR. In contrast, that number was 945 for the six months prior to GDPR~\cite{pre-gdpr-numbers}.

\section{Designing for Compliance}
\label{sec-design}
We analyze GDPR articles, both individually and collectively, from a database perspective. The goal of this section is to distill our analysis into attributes and actions that correspond to database systems. We identify three overarching themes: how personal data is to be represented, how personal data is to be protected, and what interfaces need to be designed for personal data. These three collectively determine how GDPR impacts database systems.

\subsection{Characterizing Personal Data}
\label{sec-design-metadata}
GDPR defines personal data to be any information concerning a natural person. As such, it includes both personally identifiable information like credit card numbers as well as information that may not be unique to a person, say search terms sent to Google. Such interpretation vastly increases the proportion of data that comes under GDPR purview. Also, by not restricting the applicability of GDPR to any particular domain like health and education as in the case of HIPAA~\cite{hipaa} and FERPA~\cite{ferpa} respectively, GDPR brings in virtually all industries under its foray. 

Next, to govern the lifecycle of personal data, GDPR introduces several behavioral characteristics associated with it; we call these \emph{GDPR metadata}. This constitutes a big departure from the evolution of data processing systems, which have typically viewed data as a helper resource that could be fungibly used by software programs to achieve their goals. We discover that, when taken collectively, these metadata attributes convert personal data from an inert entity to \emph{a dynamic entity} that possesses its own purpose, objections, time-to-live etc., such that it can no longer be used as a fungible commodity. Below, we list the seven metadata attributes that must be stored along with every piece of personal data\footnote{\small Though it may be possible for controllers to reduce this overhead by reusing the same metadata across groups of related items like GPS traces of a given person, the metadata granularity cannot be increased to per-person or per-service level.}.

\begin{enumerate}[leftmargin=4mm, parsep=1mm, topsep=1mm]

\item \textbf{Purpose.} \gdpr5(1b) states that personal data should only be collected for specific and explicit purposes, and not further processed in a manner incompatible with those purposes. Also, the recent Google case~\cite{google-purpose-bundling} has established that GDPR explicitly prohibits any purpose bundling.

\item \textbf{Time to live.} Given the value of personal data, the longstanding practice in computing has been to preserve them indefinitely (or at least till they are economically viable). However, \gdpr5(1e) mandates that personal data shall be kept for no longer than necessary for the purposes for which it was collected. In addition, \gdpr13(2a) requires the controller to provide this TTL value to the customer at the time of data collection. 

\item \textbf{Objections.} \gdpr21 grants users a right to object, at any time, to any personal data being used for any purposes. This right is broadly construed, and a controller has to demonstrate compelling legitimate grounds to override it. This property, essentially sets up a blacklist for every personal data item.

\item \textbf{Audit trail.} \gdpr30 requires controllers and processors to maintain records of their processing activities. Then, \gdpr33(3a) requires that in the event of a data breach, the controller shall report the number of customers affected as well as details about their records exposed. In light of these requirements, cloud providers including Amazon~\cite{aws-gdpr-list} have started supporting fine-grained per-item access monitoring. This would create an audit trail for every personal data item.

\item \textbf{Origin and sharing.} Every personal data item should have an origin i.e., how it was originally procured, and sharing information i.e., external entities with which it has been shared (\gdpr13, 14). The data trail set up by these articles should enable customers to track their personal data in secondary markets.

\item \textbf{Automated decision-making.} This concerns the emerging use of algorithmic decision-making. \gdpr15(1) grants customers a right to seek information on which of their personal data was used in automated decision-making. Conversely, \gdpr22 allows them to request that their personal data be not used for automated decision-making. 

\item \textbf{Associated person.} \gdpr15 enables users to ask for all the personal data that concern them along with all the associated GDPR metadata. As such, it is imperative to store the identification of the person to whom it concerns.

\end{enumerate}

\vheading{Impact on Database System Design.}
We call our observation that every personal data item should be associated with a set of GDPR metadata properties as \textbf{\textit{metadata explosion}}. This has significant consequences in both control- and data-path operations of database systems. First, having to store metadata along with the data increases the overall storage space. Second, having to validate each access (for purpose etc.,) and having to update after each access (for audit etc.,), increases the access latency for personal data. While optimizations like metadata normalization or metadata-aware caching could help minimize this overhead, the resulting overhead would still be significant.

%--------------------------------------------------------------------
% GDPR feature Table
%--------------------------------------------------------------------
\begin{table*}[t]
\caption{\textbf{The table maps the requirements of key GDPR articles into database system attributes and actions. This provides a blueprint for designing new database systems as well as retrofitting the current systems into GDPR compliance.}}
\label{fig:regulation-table}
\makebox[1\textwidth][c]{
\begin{minipage}[b]{1\textwidth}
\centering
{\renewcommand{\arraystretch}{1.2}
\begin{tabular}{@{}l l l | l l@{}}
\toprule[1.2pt]
\multirow{2}{*}{\makecell[c]{\bf No}} & \multirow{2}{*}{\makecell[c]{\bf GDPR article/clause}} & \multirow{2}{*}{\makecell[c]{\bf What they regulate}} & \multicolumn{2}{l}{\bf Impact on database systems} \\
 & & & \makecell[l]{Attributes} & \makecell[l]{Actions} \\
\toprule[1.1pt]
{5} & \textsc{Purpose limitation} & Collect data for explicit purposes & Purpose & Metadata indexing \\ \hline
{5} & \textsc{Storage limitation} & Do not store data indefinitely & TTL & Timely deletion \\ \hline
\makecell[l]{13\\ 14} & \textsc{\makecell[l]{Information to be provided [...]}} & \makecell[l]{Inform customers about all the GDPR\\ metadata associated with their data} & \makecell[l]{Purpose, TTL,\\ Origin, Sharing} & Metadata indexing \\ \hline
{15} & \textsc{Right of access by users} & Allow customers to access all their data & Person id & Metadata indexing \\ \hline
{17} & \textsc{Right to be forgotten} & Allow customers to erasure their data & TTL & Timely deletion \\ \hline
{21} & \textsc{Right to object} & Do not use data for any objected reasons & Objections & Metadata indexing \\ \hline
{22} & \textsc{\makecell[l]{Automated individual decision-making}} & \makecell[l]{Allow customers to withdraw from\\ fully algorithmic decision-making} & \makecell[l]{Automated\\ decisions} & Metadata indexing \\ \hline
{25} & \textsc{\makecell[l]{Data protection by design and default}} & Safeguard and restrict access to data & --- & Access control \\ \hline
{28} & \textsc{Processor} & \makecell[l]{Do not grant unlimited access to data} & --- & \makecell[l]{Access control} \\ \hline 
{30} & \textsc{Records of processing activity} & Audit all operations on personal data & Audit trail & Monitor and log \\ \hline
{32} & \textsc{Security of processing} & Implement appropriate data security & --- & Encryption \\ \hline
\makecell[l]{33} & \textsc{\makecell[l]{Notification of personal data breach}} & Share audit trails from affected systems & Audit trail & Monitor and log \\
\bottomrule[1.2pt]
\end{tabular}
}
\end{minipage}}
\end{table*}
%--------------------------------------------------------------------

\subsection{Protecting Personal Data}
GDPR declares (in \gdpr24) that those who collect and process personal data are solely responsible for its privacy and protection. Thus, it not only mandates the controllers and processors to proactively implement security measures, but also imposes the burden of proving compliance (in \gdpr5(2)) on them. Based on our analysis of GDPR, we identify five security-centric features that must be supported in the database system. 

\begin{enumerate}[leftmargin=4mm, parsep=1mm, topsep=1mm]

\item \textbf{Timely Deletion.} In addition to \gdpr5(1e) that requires every personal data item to have an expiry date, \gdpr17 grants customers the right to request erasure of their personal data at any time. Thus, datastores must have mechanisms to allow timely deletion of possibly large amounts of data.

\item \textbf{Monitoring and Logging.} As per \gdpr30 and \gdpr33(3a), the database system needs to monitor its operations in both data path (i.e., read or write) and control path (say, changes to access control), so that compliance can be established upon request by a regulator, or relevant information be shared with regulators and customers in the event of data breaches.
 
\item \textbf{Indexing via Metadata.} Ability to access groups of data based on one or more metadata fields is essential. For example, controllers needing to modify access control (\gdpr25(2)) against a given customer's data; \gdpr28(3c) allowing processors to access only those personal data for which they have requisite access and valid purpose; \gdpr15-18, 20-22 granting customers the right to act on their personal data in a collective manner (deleting, porting, downloading etc.,); finally, \gdpr31 allowing regulators to seek access to metadata belonging to affected customers.

\item \textbf{Encryption.} \gdpr32 requires controllers to implement encryption on personal data, both at rest and in transit. While pseudo-nymization may help reduce the scope and size of data needing encryption, it is still required of the datastore. 

\item \textbf{Access Control.} \gdpr25(2) calls on the controller to ensure that by default, personal data are not made accessible to an indefinite number of entities. So, to limit access to personal data based on established purposes, for permitted entities, and for a predefined duration of time, the datastore needs an access control that is fine-grained and dynamic.

\end{enumerate}

\vheading{Impact on Database System Design.}
GDPR's goal of \emph{data protection by design and by default} sits at odd with the traditional system design goals of optimizing for cost, performance, and reliability. While our analysis identified a set of just five security features, we note that modern database systems have not evolved to support these features efficiently. Thus, a fully-compliant system would likely experience significant performance degradation.

\subsection{Interfacing with Personal Data}
\label{sec-design-queries}

GDPR defines four distinct entities---controller, customer, processor, and regulator---that interface with the database systems (shown in Figure~\ref{fig:gdpr-dataflow}). Then, its articles collectively describe the control- and data-path operations that each of these entities are allowed to perform on the database system. We refer to this set of operations as \emph{GDPR queries}, and group them into seven categories:

\begin{itemize}[leftmargin=4mm, parsep=1mm, topsep=1mm]
\item{{\small \texttt{\textbf{CREATE-RECORD}}} to allow controllers to insert a record containing personal data with its associated metadata (\gdpr24).}
\item{{\small\texttt{\textbf{DELETE-RECORD-BY-}\{\textbf{KEY}$|$\textbf{PUR}$|$\textbf{TTL}$|$\textbf{USR}\}}} to allow customers to request erasure of a particular record (\gdpr17); to allow controllers to delete records corresponding to a completed purpose (\gdpr5.1b), to purge expired records (\gdpr5.1e), and to clean up all records of a particular customer.}
\item{{\small \texttt{\textbf{READ-DATA-BY-}\{\textbf{KEY}$|$\textbf{PUR}$|$\textbf{USR}$|$\textbf{OBJ}$|$\textbf{DEC}\}}} to allow processors to access individual data items or those matching a given purpose (\gdpr28); to let customers extract all their data (\gdpr20); to allow processors to get data that do not object to specific usage (\gdpr21.3) or to automated decision-making (\gdpr22).} 
\item{{\small \texttt{\textbf{READ-METADATA-BY-}\{\textbf{KEY}$|$\textbf{USR}$|$\textbf{SHR}\}}} to allow customers to find out metadata associated with their data (\gdpr15); to assist regulators to perform user-specific investigations, and investigations into third-party sharing (\gdpr13.1).}
\item{{\small \texttt{\textbf{UPDATE-DATA-BY-KEY}}} to allow customers to rectify inaccuracies in personal data (\gdpr16).}
\item{{\small \texttt{\textbf{UPDATE-METADATA-BY-\{\textbf{KEY}$|$\textbf{PUR}$|$\textbf{USR}\}}}} to allow customers to change their objections (\gdpr18.1) or alter previous consents (\gdpr7.3); to allow processors to register the use of given personal data for automated decision making (\gdpr22.3); to enable controllers to update access lists and third-party sharing information for groups of data (\gdpr13.3).}
\item{{\small \texttt{\textbf{GET-SYSTEM-}\{\textbf{LOGS}$|$\textbf{FEATURES}\}}} to enable regulators to investigate system logs based on time ranges (\gdpr33, 34), and to identify supported security capabilities (\gdpr24,25).}
\end{itemize}

\vheading{Impact on Database System Design.}
When taken in totality, GDPR queries may resemble traditional workload, but it would be remiss to ignore two significant differences: (i) there is a heavy skew of metadata-based operations, and (ii) there is a need to enforce restrictions on who could perform which operations under what conditions. These observations make it impractical to store GDPR metadata away from the personal data (say, on backup storage to save money), which in turn may affect system optimizations like caching and prefetching (since the results, and even the ability to execute a query are conditional on several metadata factors).

\subsection{Summary} 

Table--\ref{fig:regulation-table} summarizes our analysis of GDPR. We identify three significant changes needed to achieve GDPR compliance: ability to handle \emph{metadata explosion}, ability to \emph{protect data by design and by default}, and ability to support \emph{GDPR queries}. While it is clear that these changes will affect the design and operation of all contemporary database systems, we lack systematic approaches to gauge the magnitude of changes required and its associated performance impact. Towards solving these challenges, we design \emph{GDPRbench}, a functional benchmark for GDPR-compliant database systems (in \sref{sec-gdprbench}), and present a case study of retrofitting two popular databases into GDPR compliance (in \sref{sec-evaluation}).

\section{GDPRbench}
\label{sec-gdprbench}

\emph{GDPRbench} is an open-source benchmark to assess the GDPR compliance of database systems. It aims to provide quantifiable ground truth concerning correctness and performance under GDPR. In the rest of this section, we motivate the need for GDPRbench, and then present its design and implementation.

\subsection{Why (a New) Benchmark?}

As our analysis in \sref{sec-design} reveals, GDPR significantly affects the design and operation of datastores that hold personal data. However, existing benchmarks like TPC and YCSB do not recognize the abstraction of personal data: its characteristics, storage restrictions, or interfacing requirements. This is particularly troublesome given the diversity of stakeholders and their conflicting interests. For example, companies may prefer a \emph{minimal compliance} that avoids legal troubles without incurring much performance overhead or modifications to their systems. On the other hand, customers may want to see a \emph{strict compliance} that prioritizes their privacy rights over technological and business concerns of controllers. Finally, regulators need to arbitrate this customer-controller tussle in a fast-evolving technology world. Thus, having objective means of quantifying GDPR compliance is essential.

A rigorous framework would allow system designers to compare and contrast the GDPR implications of their design choices, as well as enable service providers to better calibrate their offerings. For example, many cloud providers currently report the GDPR compliance of their services in a coarse yes-no format~\cite{aws-gdpr-list}, making it difficult for regulators and customers to assess either the compliance levels or performance impact. Finally, many governments around the world are actively drafting privacy regulations. For instance, India's ongoing Personal Data Protection bill~\cite{india-privacy-bill}, and California'a Consumer Privacy Act (CCPA)~\cite{ccpa}. This benchmark provides a methodical way to study the efficacy of GDPR regulations, and then adopt suitable parts of this law. 

%--------------------------------------------------------------------
% Workload Table
%--------------------------------------------------------------------
\begin{figure*}[t]
\makebox[1\textwidth][c]{
\begin{minipage}[b]{0.75\textwidth}
\centering
\subfloat[Core Workloads\label{fig:gdprbench-workloads}]
{\renewcommand{\arraystretch}{1.1}
\begin{tabular}{@{}l l l c l@{}}
\toprule[1.2pt]
\multirow{2}{*}{\bf Workload} & \multirow{2}{*}{\bf Purpose} & \multirow{2}{*}{\bf Operations} & {\bf Default} & {\bf Default} \\
 & & & {\bf Weights} & {\bf Distrib.} \\
\toprule[1.1pt]
\multirow{3}{*}{Controller} & \multirow{3}{*}{\makecell[l]{Management and\\ administration of\\ personal data}} & \textsc{\makecell[l]{create-record}} & {25\%} & \multirow{3}{*}{Uniform} \\
 & & \textsc{\makecell[l]{delete-record-by-\{pur$|$ttl$|$usr\}}} & {25\%} & \\
 & & \textsc{\makecell[l]{update-metadata-by-\{pur$|$usr$|$shr\}}} & {50\%} & \\ \hline
\multirow{5}{*}{Customer} & \multirow{5}{*}{\makecell[l]{Exercising\\ GDPR rights}} & \textsc{{read-data-by-usr}} & {20\%} & \multirow{5}{*}{Zipf} \\
 & & \textsc{read-metadata-by-key} & {20\%} & \\
 & & \textsc{update-data-by-key} & {20\%} & \\
 & & \textsc{update-metadata-by-key} & {20\%} & \\
 & & \textsc{delete-record-by-key} & {20\%} & \\ \hline
\multirow{2}{*}{Processor} & \multirow{2}{*}{\makecell[l]{Processing of\\ personal data}} & \textsc{\makecell[l]{read-data-by-key}} & {80\%} & Zipf \\ 
 & & \textsc{read-data-by-\{pur$|$obj$|$dec\}} & {20\%} & Uniform \\ \hline 
\multirow{3}{*}{Regulator} & \multirow{3}{*}{\makecell[l]{Investigation and\\ enforcement of\\ GDPR laws}} & \textsc{\makecell[l]{read-metadata-by-usr}} & {46\%} & \multirow{3}{*}{Zipf} \\
 & & \textsc{get-system-logs} & {31\%} & \\
 & & \textsc{verify-deletion} & {23\%} & \\
\bottomrule[1.2pt]
\end{tabular}}
\end{minipage}
\begin{minipage}[]{0.25\textwidth}
\centering
\subfloat[Architecture\label{fig:gdprbench-arch}]
{\includegraphics[width=0.7\textwidth]{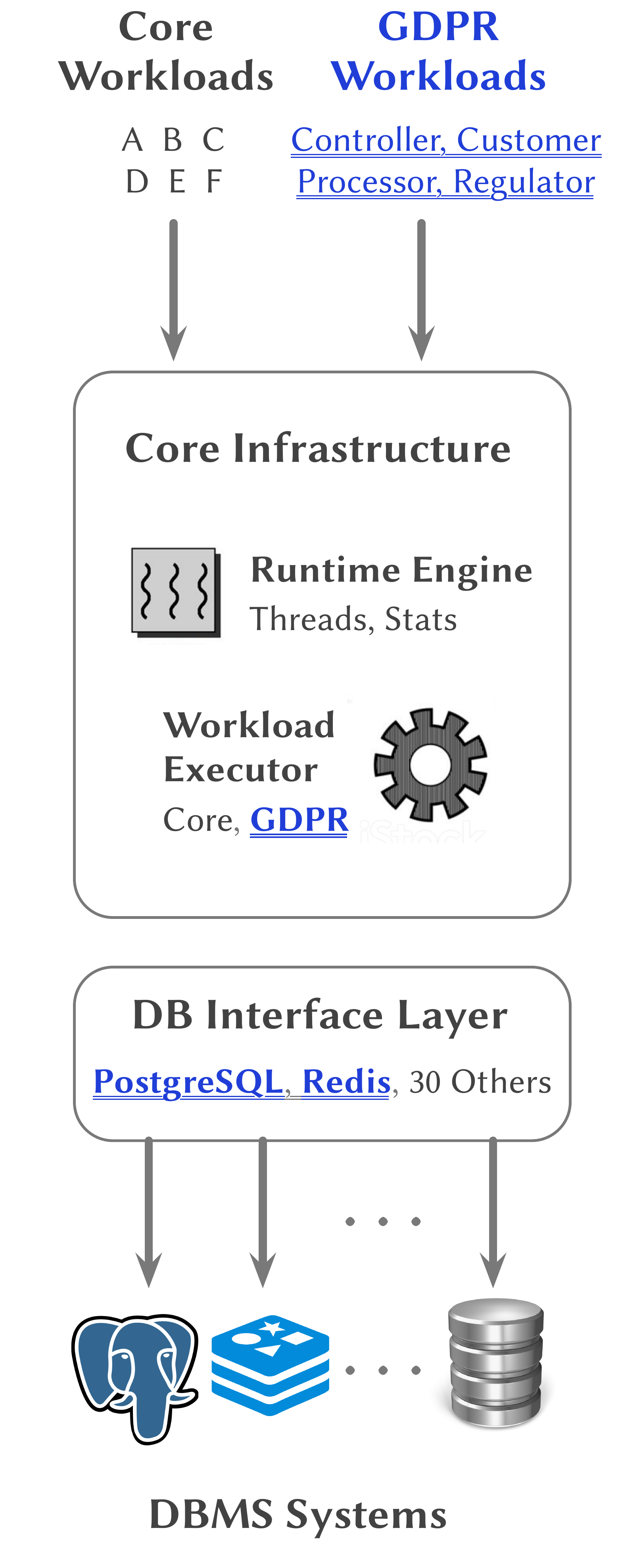}}
\end{minipage}}
\caption{\textbf {GDPRbench core workloads (a), and its architecture (b). The table describes the high-level purpose of each workload along with its composite queries and their default parameters. We select these defaults based on GDPR properties, data from EU regulators, and usage patterns from industry. The architecture diagram shows the components of YCSB that we reused in gray and our GDPR-specific extensions in {\color{blue} \underline{blue}}.}}
\vspace{-2mm}
\label{fig:workload-table}
\end{figure*}
%--------------------------------------------------------------------

\subsection{Benchmark Design}

Our approach to benchmark design is guided by two factors: insights from our GDPR analysis, and real-world data from the first year of GDPR roll out. At a high level, GDPRbench models the working of a database deployed by an organization that collects and processes personal data. Below, we describe the key elements of the benchmark design.

\vheading{4.2.1\indent Data Records}
\vspace{1mm}

Given the stringent requirements of GDPR, it is prudent to assume that personal data would be stored separately from other types of data. Thus, our benchmark exclusively focuses on personal data records. Each record takes the form \texttt {\small{\textless Key\textgreater\textless Data\textgreater}} \texttt {\small{\textless Metadata\textgreater}}, where \texttt{\small{\textless Key\textgreater}} is a variable length unique identifier, \texttt {\small{\textless Data\textgreater}} is a variable length personal data, and \texttt {\small {\textless Metadata\textgreater}} is a sequence of seven attributes, each of which has a three letter attribute name followed by a variable length list of attribute values. We enforce all the fields of the record to have ASCII characters (except for semicolon and comma, which we use as separators). We illustrate this with an example record:
 
\begin{framed}
\noindent \texttt{\small{ph-1x4b;123-456-7890;PUR=ads,2fa;TTL=7776000;
USR=neo;OBJ=$\varnothing$;DEC=$\varnothing$;SHR=$\varnothing$;SRC=first-party;}}
\end{framed}

Here, \texttt{ph-1x4b} is the unique key and \texttt{123-456-7890} is the personal data. Following these two, we have seven attributes namely purpose (PUR), time-to-live (TTL), user (USR), objections (OBJ), automated decisions (DEC), third-party sharing (SHR), and originating source (SRC). Attributes could have a single value, a list of values, or $\varnothing$. While the benchmark defines default lengths and values for all these fields, they could be modified in the configuration file to accurately represent the testing environment. Finally, we neither require nor prescribe any specific internal data layout for the personal records, and leave it up to the individual databases to organize them in the most performant way.

\vheading{4.2.2\indent Workloads}
\label{sec-gdprbench-workloads}
\vspace{1mm}

We define four workloads that correspond to the core entities of GDPR: controller, customer, processor and regulator. We compose the workloads using the queries outlined in \sref{sec-design-queries}, and concerning only the flow of personal data and its associated metadata (denoted in Figure--\ref{fig:gdpr-dataflow} by thick and dotted lines respectively). Then, we glean over usage patterns and traces from the real-world to accurately calibrate the proportion of these queries and the distribution of the data records they act on. However, since GDPR is only a year old, the availability of said data in the public domain is somewhat limited. In situations where no real data is available, we make reasonable assumptions in composing the workloads. The resulting GDPRbench workloads are summarized in Table--\ref{fig:workload-table}, and described in detail below. While GDPRbench runs these workloads in its default configuration, we make it possible to update or replace them with custom workloads, when necessary.

\vheading{Controller.}
The controller workload consists of three categories of operations: (i) creation of records, (ii) timely deletion of records, and (iii) updates to GDPR metadata towards managing access control, categorization, third-party sharing, and location management. While the controller is also responsible for the security and reliability of the underlying storage system, we expect these to be infrequent, non real-time operations and thus, do not include them in our queries.

To determine the frequency and distribution of operations, we rely on three GDPR properties: first, in a steady state, the number of records created must match the number of records deleted (since \gdpr5.1 mandates that all personal data must have an expiry date); next, a valid sequence of operation for each record should always be create, updates, and delete in that order; lastly, the controller queries should follow a uniform distribution (since \gdpr5.1c prevents the controller from collecting any data that are not necessary or useful). We set the update queries to occur twice as frequently as creates and deletes.

\vheading{Customer.}
This represents the key rights that customers exercise while interfacing with the datastore: (i) the right to delete any of their data, (ii) the right to extract and port all their data, (iii) the right to rectify personal data, and finally (iv) the right to access and update the metadata associated with a given personal data. 

To determine the frequency and distribution of customer queries, we study operational traces from Google's implementation of Right-to-be-Forgotten (RTBF)~\cite{google-forgotten}. Though GDPR has a namesake article (\gdpr17), RTBF is a distinct 2014 EU ruling that allowed individuals to request the search engines to delist URLs that contain inaccurate, irrelevant and excessively personal information from their search results. We gather three high-level takeaways from the Google report: first, they received 2.4 million requests over a span of three years at a relatively stable average of 45k monthly requests. Second, 84.5\% of all delisting requests came from individual users. Finally, the requests showed a heavy skew towards a small number of users (top 0.25\% users generated 20.8\% delisting). Based on these insights, we compose our customer workload by assigning equal weights to all query types and configuring their record selections to follow a Zipf distribution.

\vheading{Regulator.}
The role of the regulator is to investigate and enforce the GDPR laws. In case of data breaches or systematic compliance violations, the regulator would summon access to detailed system logs for the period of interest. In case of privacy rights violation of individual customers, they would seek access to the GDPR metadata associated with that particular customer. However, regulators do not need access to any personal data.

To calibrate the regulator workload, we inspect the European Data Protection Board's summary~\cite{edpb-8-months} of the first 9 months of GDPR roll out. It reports that the supervisory authorities received 206326 complaints EU-wide. Out of these, 94622 (46\%) were direct customer complaints concerning their privacy rights, 64684 (31\%) were voluntary data breach notifications from controllers, and the rest (23\%) were statutory inquiries against multinational companies, and complaints by non-government and civil rights organizations. We set the weights of regulator queries  to match the percentages from this report. Next, in line with the Google RTBF experience, we expect the rights violations and compliance complaints to follow a Zipf distribution. 

\vheading{Processor.}
The processor, working on behalf of a controller, performs a well-defined set of operations on personal data belonging to that controller. While this role is commonly external, say a cloud provider, the law also allows controllers to be processors for themselves. In either case, the processor workload is restricted to read operations on personal data.

We compose the processor workload to reflect both existing and emerging access patterns. For the former, we refer to the five representative cloud application workloads identified by YCSB, as shown in Table--\ref{tbl-ycsb}. While some operations (like updates and inserts) are not permitted for processors, their access patterns and record distributions are still relevant. For the emerging category, we rely on our GDPR analysis, which identifies access patterns conditioned on metadata attributes like purpose and objection. Since this is still an emerging category, we limit its weight to 20\%.

\vheading{4.2.3\indent Benchmark Metrics}
\vspace{1mm}

We identify three metrics that provide a foundational characterization of a database's GDPR compliance: correctness against GDPR workloads, time taken to respond to GDPR queries, and the storage space overhead.

\vheading{Correctness.}
We define correctness as the percentage of query responses that match the results expected by the benchmark. This number is computed cumulatively across all the four workloads. It is important to note that correctness as defined by GDPRbench is a \emph{necessary but not sufficient} condition for the database to be GDPR compliant. This is because GDPR compliance includes multitude of issues including data security, breach notification, prior consultation and others that cover the whole lifecycle of personal data. However, the goal of this metric is to provide a basic validation for a database's ability to implement metadata-based access control.

\vheading{Completion time.}
This metric measures the time taken to complete all the GDPR queries, and we report it separately for each workload. For majority of GDPR workloads, completion time is more relevant than the traditional metrics like latency. This is because GDPR operations embody the rights and responsibilities of the involved actors, and thus, their utility is reliant upon completing the operation (and not merely starting them). This is also reflective of the real world, where completion time gets reported more prominently than any other metric. For e.g., Google cloud guarantees that any request to delete a customer's personal data would be completed within 180 days.

\vheading{Space overhead.}
It is impossible for a database to comply with the regulations of GDPR without storing large volumes of metadata associated with personal data (a phenomenon described in \sref{sec-design-metadata} as metadata explosion). Since the quantity of metadata overshadows that of personal data, it is an important metric to track. GDPRbench reports this metric as the ratio of total size of the database to the total size of personal data in it. Thus, by definition, it will always be a rational number \textgreater 1. As a metric, storage space overhead is complementary to completion time since optimizing for one will likely worsen the other. For example, database applications can reduce the storage space overhead by normalizing the metadata. However, this will increase the completion time of GDPR queries by requiring access to multiple tables. 

\subsection{Architecture and Implementation}
We build GDPRbench by adapting and extending YCSB. This choice was driven by two factors. First, YCSB is an open-source benchmark with a modular design, making it easy to reuse its components and to build new ones on top of it. Second, it is a modern benchmark (released in 2010) and has a widespread community adoption with support for 30+ database and storage systems. 

\vheading{Benchmark architecture.} Figure--\ref{fig:gdprbench-arch} shows the core infrastructure components of YCSB (in {\color{gray} gray}), and our modifications and extensions (in {\color{blue} \underline{blue}}). Like the core workloads of YCSB, we create new GDPR workloads that describe the GDPR queries and proportions for GDPR roles (in Table--\ref{fig:workload-table}). Inside YCSB core infrastructure, we modify the workload engine to parse the GDPR queries and translate them to generic database operations. Note that we reuse the YCSB runtime engine that manages threads and statistics. All of our core infrastructure changes were done in $\sim$1300 LoC.

\vheading{Database Clients.} These modules translate the generic storage queries into native APIs that could be understood by the given database. They are also useful for implementing any missing features or abstractions as well as converting input data into formats suitable for database ingestion. Though YCSB already has client stubs for 30+ database systems, the new requirements of GDPRbench meant that we had to re-implement parts of Redis and JDBC stubs. In our client stubs, we retained the native data representation of the target databases: key-value format for Redis, and table format for the other two RDBMS. Our Redis client includes support for new GDPR queries, and a metadata-based access control. We implemented these changes in $\sim$400 lines of Java code.

\vheading{Extensions and Configurability.} GDPRbench retains the same level of extensibility and configurability as YCSB. For example, adding support for a new database simply requires adding a new client stub. Similarly, its configuration file allows modifying the workload characteristics, runtime parameters, and scale of data to better suit the testing environment.

%--------------------------------------------------------------------
% GDPR roles and data types 
%--------------------------------------------------------------------
%\begin{figure}[t]
%\centering
%\includegraphics[width=0.2\textwidth]{figures/gdprbench.pdf}
%\caption{\emph{GDPRbench architecture with YCSB components in gray and our extensions in {\color{blue} \underline{blue}}.}}
%\label{fig:gdprbench-implementation}
%\vspace{-2mm}
%\end{figure}
%--------------------------------------------------------------------

\section{GDPR-Compliant DBMS}
\label{sec-redis-postgres}

Our goal is to present a case study of retrofitting current generation systems to operate in a GDPR world. Accordingly, we select three widely used database systems: Redis~\cite{redis}, an in-memory NoSQL store, PostgreSQL~\cite{postgresql}, a fully featured RDBMS, and an commercial enterprise-grade RDBMS that we call System-C\footnote{\small{since we do not have the necessary licence to publicly share the benchmark results of this system, we anonymize its name.}}. This choice is representative of database organization (SQL vs. NoSQL), design philosophies (fully featured vs. minimalist), development methodology (open-source vs. commercial), and deployment environments. In turn, this diversity helps us generalize our findings with greater confidence. Our effort to transform Redis, PostgreSQL, and System-C into GDPR compliance is largely guided by the recommendations in their official blogs~\cite{gdpr-redis, postgresql}, and other experiences from real-world deployments. While Table-\ref{tbl-feature-matrix} succinctly represents the native support levels for GDPR security features in these three systems, we describe these and our modifications in detail below.

\vheading{Redis.}
From amongst the features outlined in \sref{sec-design}, Redis fully supports monitoring; partially supports timely deletion; but offers no native support for encryption, access control, and metadata indexing. In lieu of natively extending Redis' limited security model, we incorporate third-party modules for encryption. For data at rest, we use the Linux Unified Key Setup (LUKS)~\cite{luks}, and for data in transit, we set up transport layer security (TLS) using Stunnel~\cite{stunnel}. We defer access control to DBMS applications, and in our case, we extend the Redis client in GDPRbench to enforce metadata-based access rights. Next, while Redis offers several mechanisms to generate audit logs, we determine that piggybacking on append-only-file ({\tt AOF}) results in the least overhead. However, since {\tt AOF} only records the operations that modify Redis' state, we update its internal logic to log all interactions including reads and scans.  

%--------------------------------------
% Baseline YCSB for Redis and Postgres 
%--------------------------------------
\begin{table}[t]
\caption{\textbf{Native support for GDPR security features. Partial support is when the native feature had to be augmented with third-party libraries and/or code changes to meet GDPR requirements.}}
\label{tbl-feature-matrix}
\makebox[0.45\textwidth][c]{
\begin{minipage}[b]{0.45\textwidth}
  \centering
  \renewcommand{\arraystretch}{1.2}
  \begin{tabular}{@{}llll@{}}
    \toprule[1.2pt]
    & {Redis} & {PostgreSQL} & {System-C} \\
    \midrule
    TTL & Partial & No & No \\
    Encryption & No & Partial & Full \\
    Auditing & Full & Full & Full \\
    Metadata indexing & No & Full & Full \\
    Access control & No & Full & Full \\
    \bottomrule[1.2pt]
  \end{tabular}
\end{minipage}}
\end{table}
%--------------------------------------

%--------------------------------------------------------------------
% Evaluation figure-1
%--------------------------------------------------------------------
\begin{figure}[t]
\centering
\subfloat[Redis TTL delay\label{fig:delay-expiry}]
{\includegraphics[width=0.3\textwidth]{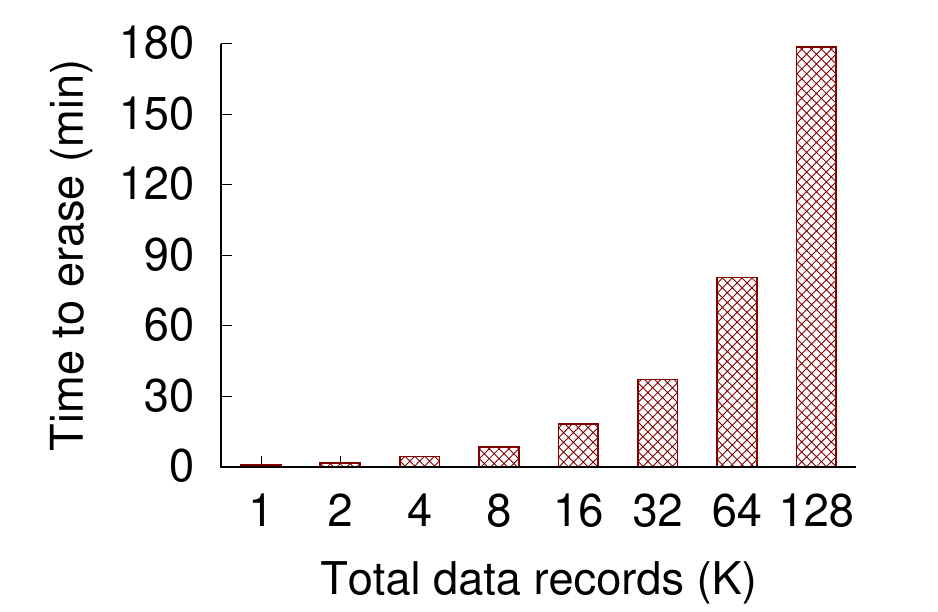}}
\subfloat[Postgres Indices\label{fig:postgres-secondary-index}]
{\includegraphics[width=0.18\textwidth]{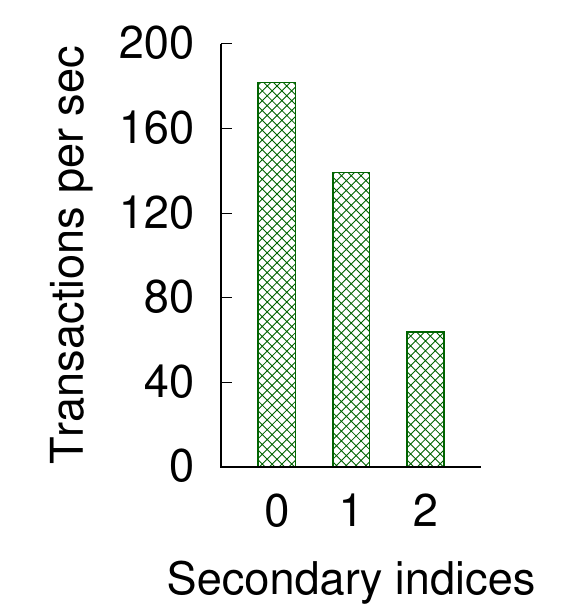}}
\caption{\textbf{Microbenchmark: (a) Redis' delay in erasing the expired keys
beyond their TTL as DB size is increased, and (b) PostgreSQL's performance worsens 
significantly as secondary indices are introduced.}}
%\vspace{-4mm}
\label{fig:microbenchmarks}
\end{figure}
%--------------------------------------------------------------------

%--------------------------------------------------------------------
% Evaluation figure-1
%--------------------------------------------------------------------
%\begin{figure}[t]
%\centering
%{\includegraphics[width=0.45\textwidth]{graphs/postgres-index-delay/indexing.pdf}}
%\caption{\emph{PostgreSQL's performance worsens significantly as secondary indices are introduced.}}
%\label{fig:postgres-secondary-index}
%\end{figure}
%--------------------------------------------------------------------

Finally, though Redis offers TTL natively, it suffers from indeterminism as it is implemented via a lazy probabilistic algorithm: once every 100ms, it samples 20 random keys from the set of keys with expire flag set; if any of these twenty have expired, they are actively deleted; if less than 5 keys got deleted, then wait till the next iteration, else repeat the loop immediately. Thus, as percentage of keys with associated expire increases, the probability of their timely deletion decreases. To quantify this delay in erasure, we populate Redis with keys having expiry times. The time-to-live values are set up such that 20\% of the keys will expire in short-term (5 minutes) and 80\% in the long-term (5 days). Figure--~\ref{fig:delay-expiry} then shows the time Redis took to completely erase the short-term keys after 5 minutes have elapsed. As expected, the time to erasure increases with the database size. For example, when there are 128k keys, clean up of expired keys ($\sim$25k of them) took nearly 3 hours. To support a stricter compliance, we modify Redis to iterate through the entire list of keys with associated {\tt EXPIRE}. Then, we re-run the same experiment to verify that all the expired keys are erased within sub-second latency for sizes of up to 1M keys.

%---------------------------
% YCSB workload description 
%---------------------------
\begin{table}[t]
\caption{\textbf{YCSB workload patterns}}
\label{tbl-ycsb}
\makebox[0.5\textwidth][c]{
\begin{minipage}[b]{0.5\textwidth}
  \centering
  \renewcommand{\arraystretch}{1.2}
  \begin{tabular}{@{}llll@{}}
    \toprule[1.2pt]
    \multicolumn{2}{@{}l}{Workload} & {Operation} & {Application} \\
    \midrule
    Load & 100\% & Insert & Bulk DB insert \\
    A & 50/50\% & Read/Update & Session store \\
    B & 95/5\% & Read/Update & Photo tagging \\
    C & 100\% & Read & User profile cache \\
    D & 95/5\% & Read/Insert & User status update \\
    E & 95/5\% & Scan/Insert & Threaded conversation \\
    F & 100\% & Read-Modify-Write & User activity record \\
    \bottomrule[1.2pt]
  \end{tabular}
\end{minipage}}
%\vspace{-5mm}
\end{table}
%--------------------------------------

\vheading{PostgreSQL.}
As a feature-rich RDBMS, PostgreSQL offers full native support to three of the five GDPR features. For encryption of data at rest, PostgreSQL does not natively support column/file-level encryption, so we set up LUKS externally. For encryption of data in transit, we setup SSL in verify-CA mode. Logging is enabled by using the built-in {\tt csvlog} in conjunction with row-level security policies that record query responses. Next, we create metadata indexing via the built-in secondary indices. As with Redis, we enforce metadata-based access control in the external client of GDPRbench. Finally, since PostgreSQL does not offer native support for time-based expiry of rows, we modify the {\texttt INSERT} queries to include an expiry timestamp and then implement a daemon that checks for expired rows periodically (currently set to 1s).

To efficiently support GDPR queries, an administrator would likely configure secondary indices for GDPR metadata. Interestingly, while PostgreSQL natively supports secondary indices, we observe that its performance begins to drop significantly when the number of such indices increases as shown in Figure--\ref{fig:postgres-secondary-index}. Using the built-in pgbench tool, we measure throughput on the Y-axis, and the number of secondary indices created on the X-axis. We run this pgbench experiment with a DB size of 15GB, a scale factor of 1000, and with 32 clients. Just introducing two secondary indices, for the widely used metadata criteria of \emph{purpose} and \emph{user-id}, reduces PostgreSQL's throughput to 33\% of the original. 

\vheading{System-C.}
Amongst the three, this offers the best level of support for GDPR. We implement TTL, the only missing feature, using the same mechanism as in PostgreSQL. For encryption, it supports Transparent Data Encryption, which encrypts the database tablespaces internally without relying on the OS/file system level encryption like LUKS. This is more secure since LUKS based encryption could allow not just the DB engine but all other applications running on the same server, an unencrypted access to the DB files. Next, we set up real-time monitoring and logging by configuring the built-in audit trail feature. Specifically, our microbenchmarks indicate that continuously streaming the generated logs to a pre-designated directory works up to 3\myx faster than saving them in the database. Finally, we configure secondary indices to improve performance for metadata-based queries of GDPRbench.

\vheading{Key Takeaways.}
\emph{Introducing GDPR compliance in Redis, PostgreSQL, and System-C was not an arduous task: Redis needed 120 lines of code changes; PostgreSQL, about 30 lines of scripting; and System-C, mostly configuration changes. We accomplished all of our modifications, configuration changes, and microbenchmarking in about two person-months. However, as our compliance effort shows, the administrators should look beyond the binary choice of whether or not a GDPR feature is supported, and analyze if a given implementation meets the expected GDPR standards.}

%--------------------------------------------------------------------
% Redis and Postgres overheads 
%--------------------------------------------------------------------
\begin{figure}[t]
%\makebox[1\textwidth][c]{
\centering
%\begin{minipage}[c]{1\textwidth}
\subfloat[Redis\label{fig:redis-ycsb-overhead}]
{\includegraphics[width=0.5\textwidth]{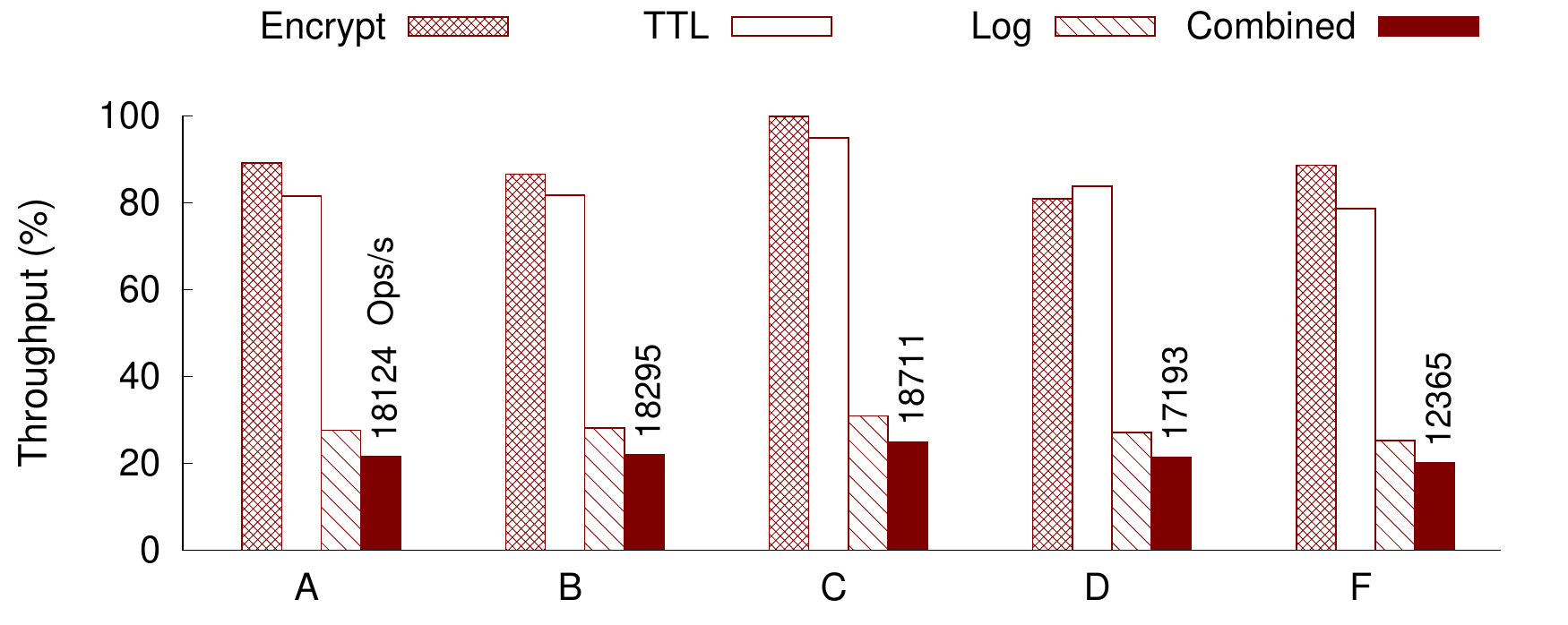}}\\
\subfloat[PostgreSQL\label{fig:postgres-ycsb-overhead}]
{\includegraphics[width=0.5\textwidth]{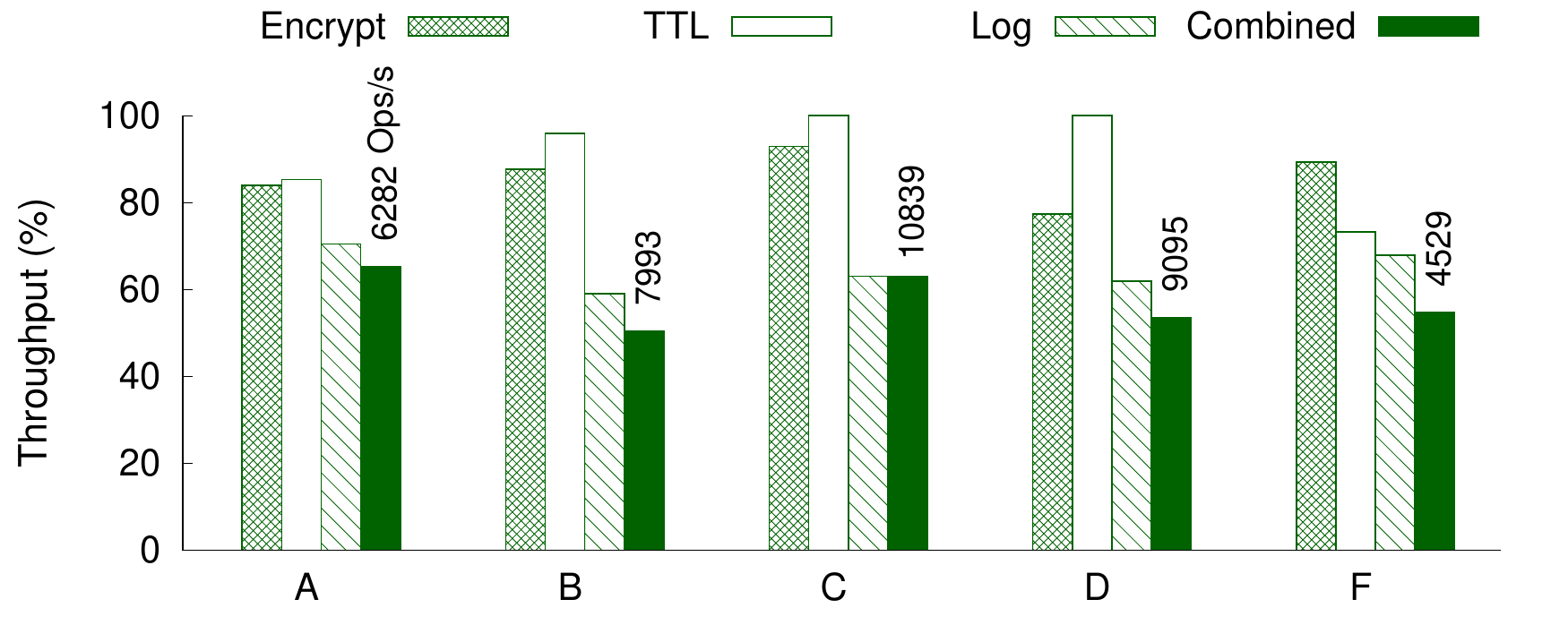}}\\
\subfloat[System-C\label{fig:oracle-ycsb-overhead}]
{\includegraphics[width=0.5\textwidth]{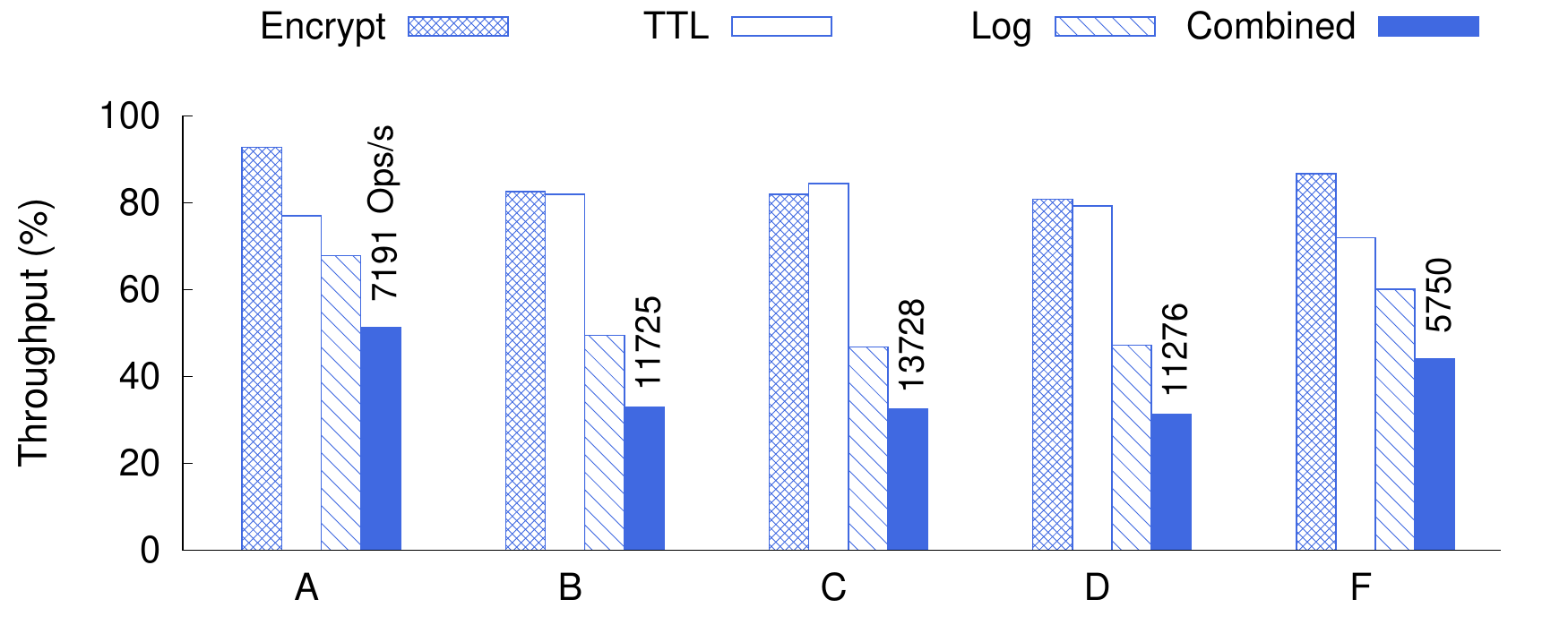}}
%\end{minipage}}
\caption{\textbf {Performance degradation after introducing GDPR features, with bars indicating performance relative to the baseline, and absolute numbers reported for the combined configuration. Our evaluation shows that when all features are enabled (solid bar), Redis experiences an overhead of 5\myx, PostgreSQL 2\myx, and System-C 2-3\myx.}}
\label{fig:redis-postgres-ycsb-overhead}
%\vspace{-4mm}
\end{figure}
%--------------------------------------------------------------------

\section{Evaluation}
\label{sec-evaluation}
We evaluate the impact of GDPR on database systems by answering the following questions:

\begin{itemize}[leftmargin=4mm, parsep=0.5mm, topsep=0.5mm]
\item{What is the overhead of GDPR features on traditional workloads? (in \sref{sec-ycsb-overhead})}
\item{How do compliant database systems perform against GDPR workloads? (in \sref{sec-gdprbench-results})}
\item{How does the scale of personal data impact performance? (in \sref{sec-gdprbench-scale})}
\end{itemize}

%--------------------------------------------------------------------
% Redis GDPRbench
%--------------------------------------------------------------------
\begin{figure*}[t]
\makebox[1\textwidth][c]{
\centering
\begin{minipage}[c]{1\textwidth}
\hspace{5mm}
\subfloat[Redis\label{fig:redis-gdprbench}]
{\includegraphics[width=0.3\textwidth]{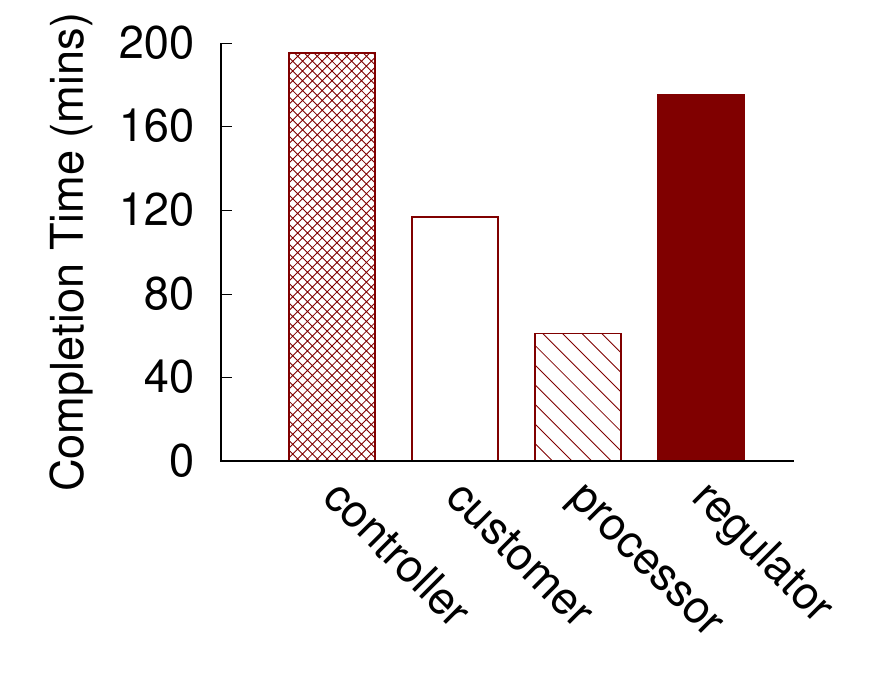}}
\hspace{3mm}
\subfloat[PostgreSQL\label{fig:postgres-gdprbench}]
{\includegraphics[width=0.3\textwidth]{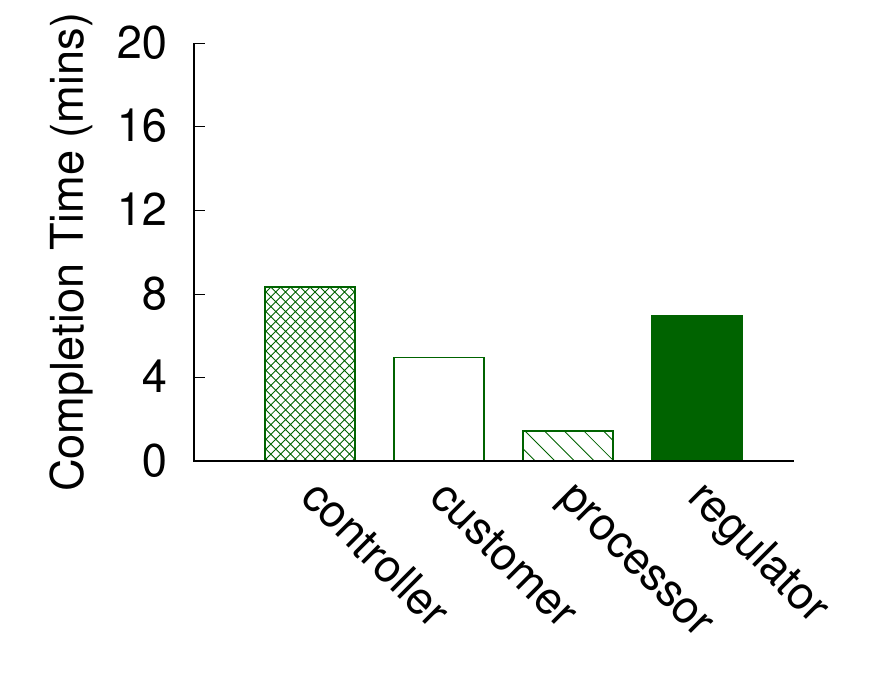}}
\hspace{3mm}
\subfloat[System-C\label{fig:oracle-gdprbench}]
{\includegraphics[width=0.3\textwidth]{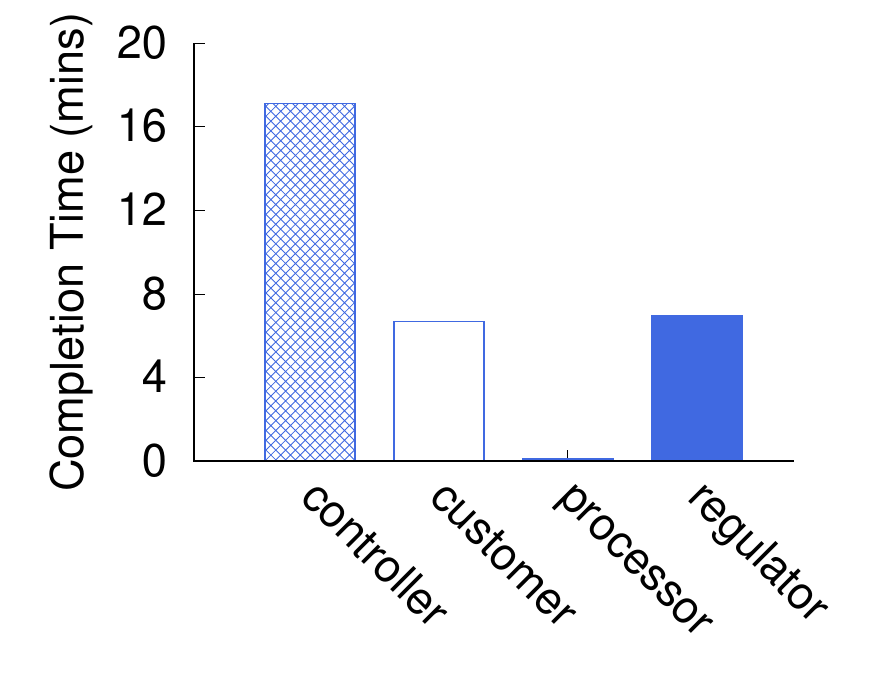}}
\end{minipage}}
\caption{\textbf {Running GDPRbench workloads on compliant versions of Redis, PostgreSQL and System-C. We see that PostgreSQL and System-C are an order of magnitude faster than Redis against GDPR workloads.}}
\label{fig:gdprbench-redis-postgres}
%\vspace{-3mm}
\end{figure*}
%--------------------------------------------------------------------

\vheading{Approach.} To answer these questions, we use the GDPR
compliant Redis, PostgreSQL, and System-C described in
Section-\ref{sec-redis-postgres}. For benchmarking against traditional
workloads, we use the industry standard Yahoo Cloud Serving
Benchmark~\cite{ycsb}, and for benchmarking against realistic GDPR
workloads, we use GDPRbench.

\vheading{Experimental setup.} We perform all our experiments on the
Chameleon Cloud~\cite{chameleon-cloud}. The database systems are run
on a dedicated Dell PowerEdge FC430 with 40-core Intel Xeon 2.3GHz
processor, 64 GB RAM, and 400GB Intel DC-series SSD. We choose Redis
v5.0 (released March 18, 2019), PostgreSQL v9.5.16 (released Feb 14,
2019), System-C (released in 2017), and YCSB 0.15.0 (released Aug 14,
2018) as our reference versions.

\vspace{2mm}
\subsection{Overheads of Compliance}
\label{sec-ycsb-overhead}

Our goal is to quantify the performance overhead of GDPR compliance by using the industry standard YCSB~\cite{ycsb}. As shown in Table--\ref{tbl-ycsb}, YCSB comprises of 6 workloads, named A through F, that represent different application patterns. For this experiment, we run YCSB with 16 threads; configure it to load 2M records and perform 2M operations in each workload category.

\vheading{Redis.}  Figure--\ref{fig:redis-ycsb-overhead} shows the
YCSB workloads on the X-axis and Redis' throughput on the Y-axis for
each of the newly introduced GDPR security features. We normalize all
the values to a baseline version of Redis that has no security or
persistence. We see that encryption reduces the throughput by
$\sim$10\%, TTL modification brings it down by $\sim$20\%, and
AoF-based logging (persisted to disk once every second) slows the
performance by $\sim$70\%. Since Redis is an in-memory datastore,
requiring it to persist AoF to the disk synchronously results in a
significant slowdown. Given that these GDPR features affect all types
of queries: read, update, and insert, the performance drops are fairly
consistent across all types of YCSB workloads. Finally, when all the
features are enabled in tandem, Redis experiences a slowdown of 5\myx.

\vheading{PostgreSQL.}
Next, Figure--\ref{fig:postgres-ycsb-overhead} shows PostgreSQL performance. PostgreSQL experiences 10-20\% degradation due to encryption and TTL checks, while logging incurs a 30-40\% overhead. When all features are enabled in conjunction, PostgreSQL slows down by a factor of 2\myx of its baseline performance. The graph also demonstrates that the effect of GDPR on PostgreSQL is not as pronounced as in the case of Redis. However, in terms of raw throughput, Redis still outperforms PostgreSQL since the former is in-memory datastore compared to the disk-based PostgreSQL.

\vheading{System-C.}  Lastly, we show System-C's performance in
Figure--\ref{fig:oracle-ycsb-overhead}. We observe that System-C's
baseline throughput is $\sim$2\myx better than PostgreSQL, and
internally, its write throughput is slightly better than its read
throughput. In terms of overheads, encryption and TTL induce 10-20\%
slowdown, but auditing causes a much steeper drop of 30-50\%. This is
surprising since System-C's audit trail natively supports the type of
extensive logging required by GDPR. Cumulatively, these features
result in a performance degradation of 2-3\myx compared to its
baseline.

\vheading{Workload E}. All three systems experienced drastic overhead
in supporting range queries. In contrast to other workloads that
finished in the order of seconds to minutes, Workload E took tens of
hours, underscoring the challenges of implementing complex queries
under GDPR settings. We omit the Workload E results as they would
require weeks to obtain; at this point, we note that implementing
range queries in an efficient manner under GDPR constraints is a
significant challenge. Note that YCSB workloads do not exercise access
control or metadata indexing, but the GDPR workloads in the next
section will incorporate these two.

\vheading{Summary.}  \emph{The security features introduced affect all
  read and write operations, resulting in reduced
  performance. Features such as logging can result in significant
  performance degradation, making GDPR compliance challenging for
  production environments.}

%--------------------------------------
% Baseline YCSB for Redis and Postgres 
%--------------------------------------
\begin{table}[t]
\caption{\textbf{Storage space overhead corresponding to Figure--\ref{fig:gdprbench-redis-postgres}. In the default configuration, GDPRbench has 25 bytes of metadata attributes for a 10 byte personal data record.}}
\label{tbl-metadata-explosion}
\makebox[0.5\textwidth][c]{
\hspace{-6mm}
\begin{minipage}[b]{0.5\textwidth}
  \centering
  \renewcommand{\arraystretch}{1.25}
  \begin{tabular}{@{}lccc@{}}
    \toprule[1.2pt]
    & \makecell[c]{Personal data\\ size (MB)} & \makecell[c]{Total DB\\ size (MB)} & \makecell[c]{Space\\ factor} \\
    \midrule
    Redis & 10 & 35 & 3.5\myx \\
    PostgreSQL & 10 & 59.5 & 5.95\myx \\
    \bottomrule[1.2pt]
  \end{tabular}
\end{minipage}}
%\vspace{-4mm}
\end{table}
%--------------------------------------

%--------------------------------------------------------------------
% Evaluation figure-1
%--------------------------------------------------------------------
%\begin{figure}[]
%\centering
%{\includegraphics[width=0.4\textwidth]{graphs/redis-vs-postgres/compare.pdf}}
%\caption{\emph{Representative throughput achieved by Redis and PostgreSQL on YCSB and GDPRbench, under identical conditions. Both systems perform 2-4 orders of magnitude worse for GDPR workloads as opposed to traditional workloads.}}
%\label{fig:redis-vs-postgres}
%\end{figure}
%--------------------------------------------------------------------

\vspace{2mm}
\subsection{GDPR Workloads}
\label{sec-gdprbench-results}

In this section, our goal is to quantify how compliant Redis,
PostgreSQL, and System-C perform against real-world GDPR workloads. A
major difference between YCSB and GDPRbench workloads is that
GDPRbench mostly consists of metadata-based queries as opposed to
primary-key based queries of YCSB. Most YCSB queries return single
record associated with one primary key. In contrast, GDPR queries such
as ``get all metadata associated with a user's records'' can return a
large number of records. As a result, GDPR queries obtain lower
throughput than YCSB queries. Cognizant of this characteristic, we
configure GDPRbench to load 100K personal records, and perform 10K
operations for each of its four workloads. Note how these numbers are
one to two orders of magnitude lower than the YCSB configuration in
Section--\ref{sec-ycsb-overhead}. We use the default proportion of
workload queries and record distributions as specified in
Table--\ref{fig:workload-table}, and run it with 8 threads.

\vheading{Redis.} 
Figure--\ref{fig:redis-gdprbench} shows Redis' completion time along Y-axis, and the GDPRbench workloads along the X-axis. As expected, the processor workload runs the fastest given its heavy skew towards non-metadata based operations. In comparison, all other workloads are 2-4\myx slower, with the controller and regulator workloads taking the longest. This is because, the customer workload simply deals with the records of a given customer, whereas the controller and regulator queries touch records across all customers. Table--\ref{tbl-metadata-explosion} benchmarks the storage overhead. In the default configuration, we see a space overhead ratio of 3.5 i.e., on average every byte of personal data inserted causes the storage size to grow by 3.5 bytes. 

\vheading{PostgreSQL.} 
Next, Figure--\ref{fig:postgres-gdprbench} shows the corresponding baseline compliance graph for PostgreSQL. Right away, we see that the completion times are an order of magnitude faster than Redis for all the workloads, while holding similar trends across the four workloads. Our profiling indicates that PostgreSQL, being an RDBMS, is better at supporting complex queries efficiently. PostgreSQL's performance is also bolstered (by $\sim$30\%) by the use of secondary indices for the metadata. However, adding these extra indices increase the storage space overhead from 3.5\myx to 5.95\myx as outlined in Table--\ref{tbl-metadata-explosion}.

\vheading{System-C.}  Figure--\ref{fig:oracle-gdprbench} shows
System-C's performance against GDPR workloads. Since GDPR-compliant
System-C came out slightly worse than PostgreSQL in our setup (as
discussed in Section--\ref{sec-ycsb-overhead}), many GDPR workloads
have taken longer to complete in System-C than PostgreSQL. However,
interestingly, since the GDPR workload sizes are smaller (10MB)
compared to the YCSB workloads (2GB), System-C's built-in query result
cache was able to significantly improve the performance of read-only
workloads such as the processor.

%--------------------------------------------------------------------
% GDPRbench scale experiments
%--------------------------------------------------------------------
\begin{figure}[t]
\centering
\subfloat[YCSB workload-C\label{fig:redis-scale-c}]
{\includegraphics[width=0.25\textwidth]{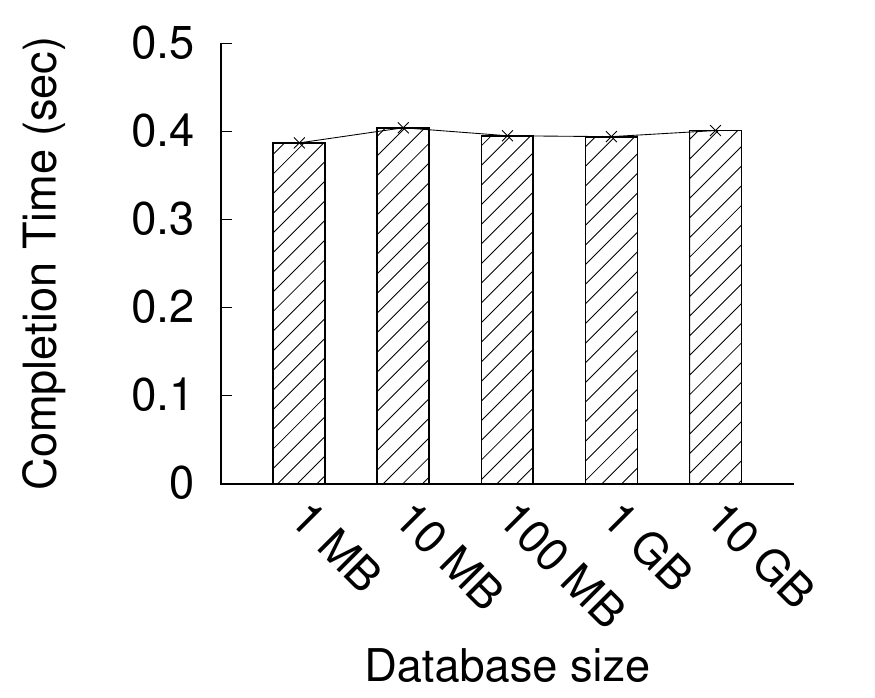}}
\subfloat[GDPRbench Customer\label{fig:redis-scale-customer}]
{\includegraphics[width=0.25\textwidth]{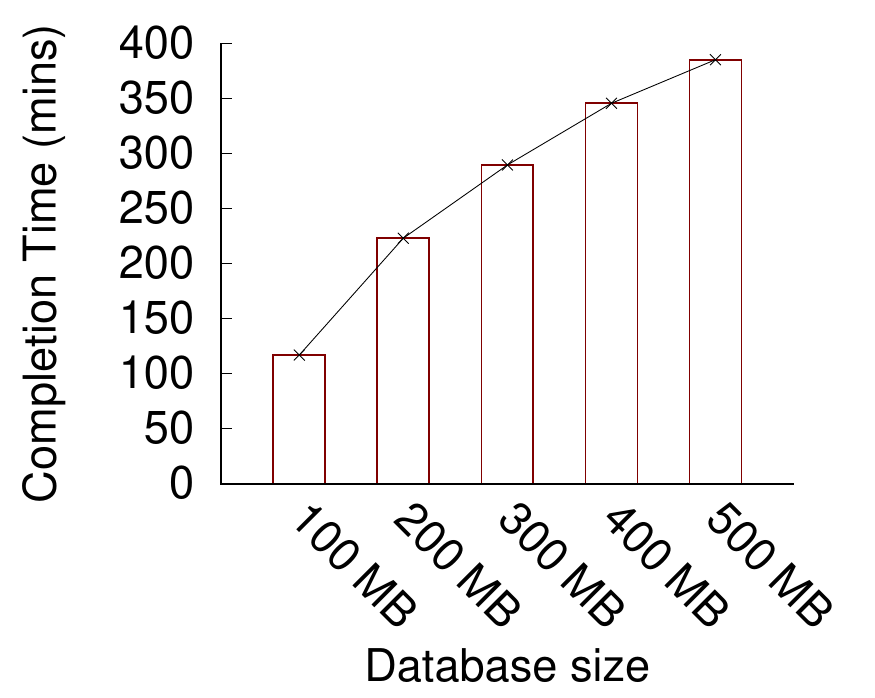}}
\caption{\textbf {Time taken by Redis to complete 10K operations as the volume of data stored in the DB increases. For the traditional workload in (a), Redis' performance is only governed by the number of operations, and thus remains virtually constant across four orders of magnitude change in DB size. However, for GDPR workload in (b), the completion time linearly increases with the DB size.}}
\label{fig:redis-scale}
\vspace{-2mm}
\end{figure}
%--------------------------------------------------------------------

\vheading{Summary.}
\emph{GDPRbench reflects the challenges of supporting GDPR workloads on retrofitted compliant systems. While all systems experience significant degradation in their performance compared to traditional workloads, our evaluation shows that feature-rich RDBMS like PostgreSQL performs better than NoSQL stores like Redis.}

\subsection{Effect of Scale}
\label{sec-gdprbench-scale}

Finally, we explore how an increase in the scale of data affects the systems. In particular, we structure this experiment to reflect a scenario where a company acquires new customers, thus increasing the volume of data in the DB. However, the data of the existing customers remain unchanged. This experiment then measures how Redis and PostgreSQL perform for queries concerning the original set of customers. We lay out experiments in two different contexts: first, when the database contains non-personal data, we run YCSB workloads; second, when the database contains personal data, we use GDPRbench customer workload. In both cases, we scale the volume of data within the database but perform the same number of operations at every scale. For both GDPR and traditional workloads, we use identical underlying hardware, same version of GDPR-compliant Redis and PostgreSQL software, and retain the same configuration as in Section--\ref{sec-ycsb-overhead}.

%--------------------------------------------------------------------
% YCSB scale experiments
%--------------------------------------------------------------------
\begin{figure}[t]
\centering
\subfloat[YCSB Workload-C\label{fig:postgres-scale-c}]
{\includegraphics[width=0.25\textwidth]{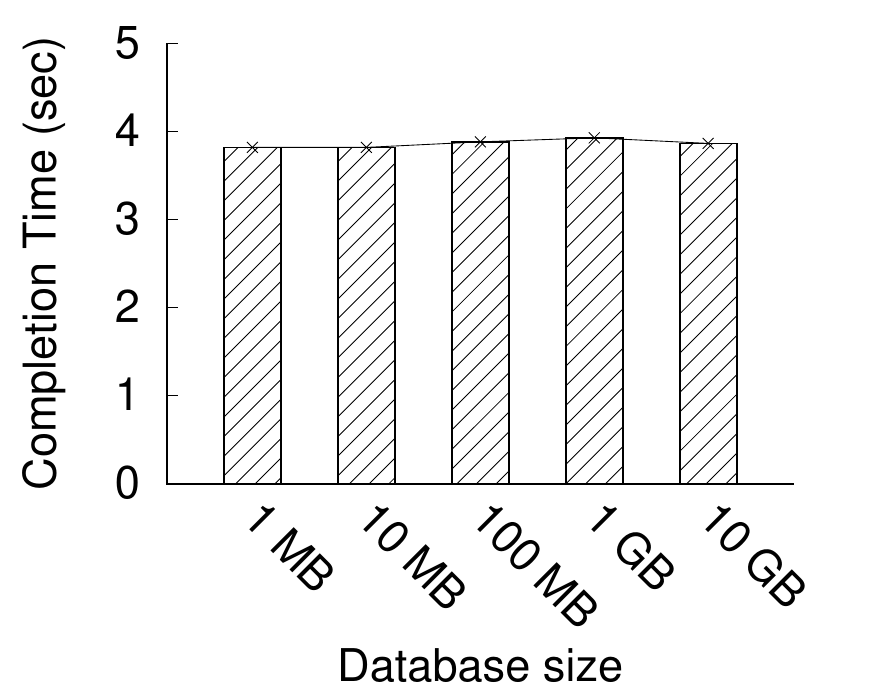}}
\subfloat[GDPRbench Customer\label{fig:postgres-scale-customer}]
{\includegraphics[width=0.25\textwidth]{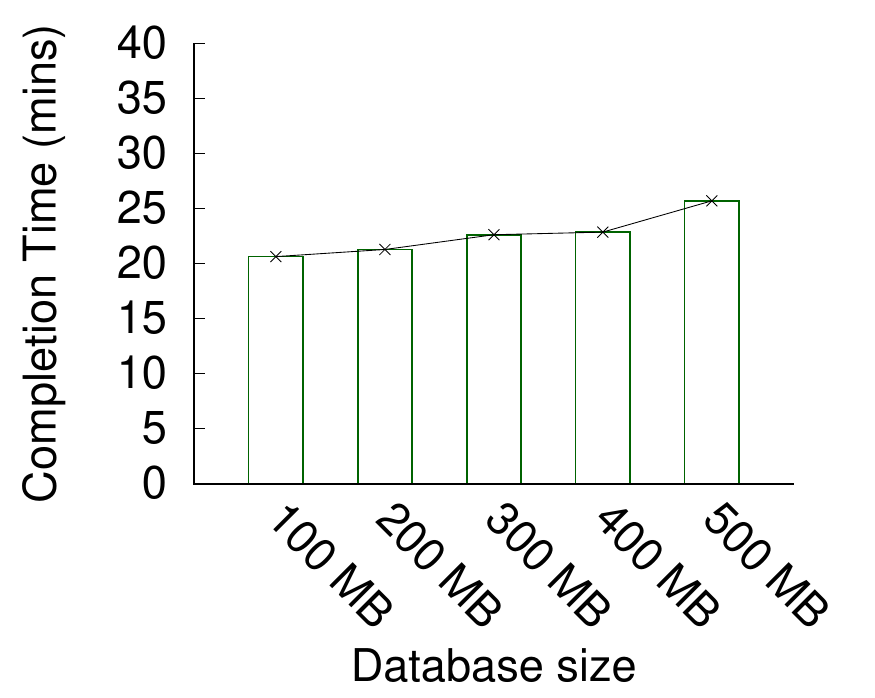}}
\caption{\textbf {Time taken by PostgreSQL to complete 10K operations as the DB size scales. Expectedly, PostgreSQL's performance remains constant for traditional workloads in (a). However, unlike in Redis (Figure-\ref{fig:redis-scale-c}), PostgreSQL's GDPR performance worsens only moderately thanks to its use of metadata indices.}}
\label{fig:postgres-scale}
\vspace{-2mm}
\end{figure}
%--------------------------------------------------------------------

\vheading{Redis.} 
We seed Redis store with 1MB worth of data and perform 10K operations using YCSB workload-C. Figure-\ref{fig:redis-scale-c} shows that Redis takes almost identical time to complete 10K operations, despite increasing the database volume by four orders of magnitude. This is not unexpected as Redis supports efficient, constant-time CRUD operations. However, when we switch from this traditional workload to a GDPR workload, Figure-\ref{fig:redis-scale-customer} paints a different picture. In this graph, we linearly increase the volume of personal data from 100 to 500MB, and we see a corresponding linear increase in the completion time. This indicates that the completion time is not only a function of the number of operations but also the size of the database. In hindsight, this is not completely unexpected as metadata based queries require O(n) access, especially in absence of secondary indices.

\vheading{PostgreSQL.} 
Next, we conduct the same scale experiment on PostgreSQL, which has support for secondary indices. While PostgreSQL's performance for YCSB (shown in Figure-\ref{fig:postgres-scale-c}) is expectedly similar to that of Redis, its response to GDPR workload (shown in Figure-\ref{fig:postgres-scale-customer}) is much better than that of Redis. While PostgreSQL is still affected by the increase in DB size, the impact on its performance is muted. Our profiling indicates that this is largely due to secondary indices speeding up metadata based queries. But as the DB size increases, the overhead of maintaining multiple secondary indices does introduce some performance degradation. 

\vheading{Summary.}  \emph{Current generation database systems do not
  scale well for GDPR workloads. PostgreSQL with metadata indexing
  fares better than Redis, but still experiences some performance
  degradation as the amount of personal data increases. }

\section{Discussion}
\label{sec-conclusion}
Our experiments and analyses identify several implications for administering GDPR-compliant database systems in the real world and research challenges emerging from it. We discuss them below. 

\vheading{Compliance may result in high performance overheads.}  Our
work demonstrates that introducing GDPR compliance into modern
database systems is straight-forward, ranging from minor code changes
in open-source systems, to simple configurations in enterprise level
systems. However, in all these case, the resulting performance
degradation of 2-5\myx (in Section-\ref{sec-ycsb-overhead}) raises
fundamental questions of compliance-efficiency tradeoffs. Database
engineers and administrators should carefully analyze the performance
implications of any compliance efforts, especially in production
environments. For instance, recommendations from cloud providers such
as Amazon Web Services~\cite{aws-gdpr-ready}, Microsoft
Azure~\cite{azure-gdpr-ready}, and Google Cloud~\cite{gce-gdpr-ready}
primarily focus on checklist of security features without much
attention to their performance implications.

\vheading{Compliant systems experience challenges at scale.}
A key takeaway from our scale experiments (in Section-\ref{sec-gdprbench-scale}) is that naive efforts at achieving GDPR compliance results in poor scalability. Increasing the volume of personal data, even by modest amounts, makes it challenging to respond to customer's GDPR rights in a timely manner, or even to comply with GDPR responsibilities in real-time. Thus, consideration for scale ought to be an important factor in any compliance effort.

Additionally, GDPR quells the notion that personal data, once collected, is largely immutable. In light of GDPR's \emph{right to be forgotten} and \emph{right to rectification}, customers are allowed to exercise much greater control over their personal data. Consequently, traditional solutions to scale problems like replication and sharding would likely incur extra overheads than before. It might be worth investigating the benefits of a GDPR co-processor. 

\vheading{Compliance is easier in RDBMS than NoSQL.}
We observe that achieving compliance is simpler and effective with RDBMSs than NoSQL stores. In our case, Redis needed two changes at the internal design level as opposed to PostgreSQL and commercial-RDBMS, which only needed configuration changes and external scripting. Even from a performance point of view, the drop is steeper in high-performant Redis as compared to the RDBMSs. We hope our findings encourage designers and maintainers of all categories of database systems to reevaluate their design choices, optimization goals, and deployment scenarios in the light of privacy regulations like GDPR. 

\vheading{GDPR is strict in principle yet flexible in practice.}
Though GDPR is clear in its high-level goals, it is intentionally vague in its technical specifications. Consider \gdpr17 that requires controllers to erase personal data upon request by the customer. It does not specify how soon after the request should the data be removed. Let us consider its implications in the real world: Google cloud, which claims GDPR-compliance, describes the deletion of customer data as an iterative process~\cite{google-deletion} that could take up to \textsc{180 days} to fully complete. Such flexibility also exists for the strength of encryption algorithms, duration for preservation of audit trails, etc. 

This flexibility in GDPR interpretation allows compliance to be treated more like a spectrum instead of a fixed target. Database engineers and administrators could use GDPRbench to explore the tradeoff between strict compliance vs. high performance. We note that compliance efforts could go beyond our choices of reusing the existing features and sticking with the default data layouts in order to improve performance. For example, GDPR metadata attributes could be shared between related groups of data, be stored in a separate table, be normalized across multiple records, or be cached in the application to reduce the storage overhead and access latency.

\vheading{Research challenges.}
Our evaluations show that trivially extending the existing mechanisms and policies to achieve compliance would result in significant performance overheads. We observe two common sources of this: (i) retrofitting new features when they do not align with the core design principles. For example, adding to Redis' minimalist security model, and (ii) using features in ways that are not intended by their designers. For example, enabling continuous auditing in a production environment. We identify three key challenges that must be addressed to achieve compliance efficiently: \emph{efficient auditing}, \emph{efficient time-based deletion}, and \emph{efficient metadata indexing}.

%Another key tussle in the design space is whether to build compliance at the level of individual infrastructure components (i.e., compute servers, and database systems) versus implementing end-to-end compliance of given regulations (i.e., implementing right-of-access in a music streaming service). Both these directions will result in different performance tradeoffs and give rise to different system interfaces. The former approach makes the effort more contained and thus, suits the cloud model better (where GDPR explicitly prohibits selling products and services that do not comply with its regulations). The latter approach provides opportunities for cross-layer optimizations (e.g., avoiding access control in multiple layers). 
%\vspace{2mm}
\section{Related Work}
\label{sec-related}

A preliminary version of this analysis appeared~\cite{gdpr-storage} in a workshop. To the best of our knowledge, this work is one of the first to analyze the impact of GDPR on database systems. While there have been a number of recent work analyzing GDPR from privacy and legal perspectives~\cite{poly-privacy-policy, cacm-gdpr-impact, counterfactual, gdpr-purpose, gdpr-www, explainable-machines, uninformed-consent, gdpr-cookies, gdpr-formal}, the database and systems communities are just beginning to get involved. DatumDB~\cite{poly-datumdb} proposes an architectural vision for a database that natively supports guaranteed deletion and consent management. Compliance by construction~\cite{poly-compliance-by-construction} envisions new database abstractions to support privacy rights. In contrast, we focus on the challenges that existing DBMS face in complying with GDPR, and design a benchmark to quantify its impact.

Orthogonal to our focus, researchers are working on implementing and analyzing individual GDPR articles end-to-end. For example, Google researchers~\cite{google-forgotten} have chronicled their experiences implementing the \emph{Right to be Forgotten} for their search service. Two groups of researchers from Oxford University analyzed~\cite{gdpr-explanation, gdpr-no-right-to-explanation} how GDPR's right to explanation impacts the design of machine learning and artificial intelligence systems. Finally, there is a wealth of blog posts that describe how to achieve GDPR compliance for popular database systems including MongoDB~\cite{gdpr-mongodb}, CockroachDB~\cite{gdpr-cockroach}, Redis~\cite{gdpr-redis}, Oracle~\cite{oracle-gdpr}, and Microsoft SQL~\cite{microsoft-sql-gdpr}.

%\vspace{2mm}
\section{Conclusion}
\label{sec-conclusion}

This work analyzes GDPR from a database systems perspective. We discover the phenomenon of metadata explosion, identify new workloads of GDPR, and design a new benchmark for quantifying GDPR compliance. We find that despite needing to implement a modest number of changes to storage systems, GDPR compliance results in significant performance overheads. Our analyses and experiments identify several implications for administering GDPR-compliant database systems in the real world. We hope that GDPRbench would be useful for customers, controllers, and regulators in interpreting the compliance level of storage systems, and helpful for database designers in understanding the compliance-performance tradeoff.

%\vspace{2mm}
\vheading{\bf Acknowledgments.}
We would like to thank our anonymous reviewers, and members of the LASR group and the Systems and Storage Lab at UT Austin for their
feedback and guidance. This work was supported by generous donations from VMware, Google, and Facebook. Vinay's work has been supported by a ReportBee Remote Fellowship. Any opinions, findings, conclusions, or recommendations expressed herein are those of the authors; these should neither be interpreted as legal advice nor as reflective of the views of their host institutions.

}

\balance
\bibliographystyle{abbrv}
\bibliography{paper}  

\end{document}